\documentclass[fleqn,10pt]{wlscirep}
\usepackage[utf8]{inputenc}
\usepackage[T1]{fontenc}
\usepackage[right]{lineno}
\linenumbers

\usepackage{color}
\usepackage{multirow,array}
\usepackage{url}

\title{Distinguishing Pairwise and Higher-Order Interactions in Coupled Oscillators from Time Series} 

\author[1]{Weiwei Su}
\author[2]{Shigefumi Hata}
\author[1,3]{Hiroshi Kori}
\author[4,5]{Hiroya Nakao}
\author[1,6*]{Ryota Kobayashi}
\affil[1]{Department of Complexity Science and Engineering, The University of Tokyo, Kashiwa, 277-0882, Japan}
\affil[2]{Graduate School of Science and Engineering, Kagoshima University, Kagoshima, 890-0065, Japan}
\affil[3]{Department of Mathematical Informatics, The University of Tokyo, Bunkyo, 113-8654, Japan}
\affil[4]{Department of Systems and Control Engineering, Institute of Science Tokyo, Meguro, 152-8552, Japan}
\affil[5]{Research Center for Autonomous Systems Materialogy, Institute of Science Tokyo, Yokohama, 226-8501, Japan}
\affil[6]{Mathematics and Informatics Center, The University of Tokyo, Bunkyo, 113-8656, Japan}

\affil[*]{r-koba@k.u-tokyo.ac.jp}

\begin{abstract}
Rhythmic phenomena, which are ubiquitous in biological systems, are typically modelled as systems of coupled limit cycle oscillators. 
Recently, there has been an increased interest in understanding the impact of higher-order interactions on the population dynamics of coupled oscillators. Meanwhile, the estimation of a mathematical model from experimental data is an essential step in understanding the dynamics of real-world complex systems. 
In coupled oscillator systems, identifying the type of interaction (e.g. pairwise or three-body) is challenging, because different interactions can exhibit similar dynamical states in experimental conditions. 
In this study, we have developed a method based on the adaptive LASSO (Least Absolute Shrinkage and Selection Operator) to infer the interactions among oscillators from time series data. 
The proposed method successfully identifies the type of interaction and estimates the probabilities of pairwise and three-body couplings.  
Through systematic analysis of synthetic datasets, we have demonstrated that our method outperforms two baseline methods, LASSO and OLS (Ordinary Least Squares), in accurately inferring the topology and strength of couplings between oscillators. 
Furthermore, the proposed method is applied to human brain network data, demonstrating its practical utility. Finally, we extend the method to more general oscillatory systems, including those exhibiting the deformation of limit cycles and those with four-body interactions. These results suggest that our method is a promising tool for identifying interaction mechanisms in oscillatory systems.  
\end{abstract}
\begin{document}
\nolinenumbers
\flushbottom
\maketitle
%
%
\thispagestyle{empty}


\section*{Introduction}

Rhythmic phenomena are present in a variety of biological systems, including the brain activity~\cite{buzsaki2004,wang2010,kobayashi2016estimation}, circadian rhythms~\cite{mohawk2011cell,yamaguchi2013mice}, heartbeats and respiration~\cite{schafer1998,lotrivc2000,kralemann2013vivo}, and animal gaits~\cite{collins1993coupled,funato2016evaluation,arai2024interlimb}. 
Mathematically, these phenomena can be modelled as systems of coupled limit cycle oscillators~\cite{strogatz2018nonlinear}.
Phase reduction theory~\cite{kuramoto1984chemical, hoppensteadt1997, nakao2016} allows us to approximate the dynamics of systems of interacting
limit cycle oscillators to coupled one-dimensional phase oscillators, also known as the phase model. 
Analysis of the phase model provides insights into the mechanisms of synchronization and cluster formation between oscillators, as well as their bifurcations. 
These theoretical studies on the phase models have elucidated key factors underlying the synchronization phenomena, including the effect of periodic external forces, and coupling between the oscillators~\cite{pikovsky2003synchronization,nakao2016}.

A complex system can be modelled as a network of interacting nodes or vertices (i.e. elements) connected by links or edges (i.e. individual interaction or couplings). However, in many systems, such as social, ecological, and biological systems, the effects of interaction between groups of three or more elements (higher-order interactions) cannot be ignored. Recently, there has been a significant amount of research on the effects of higher-order interactions between the elements on the network dynamics~\cite{battiston2020networks,bick2023higher,boccaletti2023structure}. 
In coupled oscillator systems, it has been reported that higher-order interactions can emerge from pairwise interaction, e.g., via the nonlinearity of the coupling~\cite{ashwin2016hopf,leon2019phase} and the dependence of the amplitude on the phase~\cite{bick2024higher}. In addition, higher-order interactions can arise spontaneously due to frequency resonance of oscillators~\cite{nijholt2022emergent}. 
There is also extensive research on the effects of higher-order interactions on dynamics~\cite{battiston2020networks}. 
Higher-order interactions in coupled oscillators exhibit richer dynamical properties compared to pairwise interactions, including multi-stability~\cite{tanaka2011multistable}, chaos~\cite{bick2016chaos} and explosive synchronisation~\cite{skardal2020higher}. 
Furthermore, higher-order interactions can induce dynamical states such as synchronisation, asynchronisation, and cluster states, whose bifurcation structure is mathematically distinct from pairwise interactions~\cite{skardal2020higher,leon2024higher,marui2025synchronization}. \newpage

Estimating a mathematical model from experimental data is an essential step in understanding the dynamics of complex systems in the real world.
In the context of coupled oscillator systems, previous studies have developed methods for identifying phase models from data. These methodologies include estimating the phase response curves~\cite{galan2005efficient,tsubo2007,nakae2010bayesian}, phase coupling functions~\cite{rosenblum2001detecting,kralemann2007uncovering,tokuda2007inferring,stankovski2012inference} (see also review~\cite{stankovski2017coupling,rosenblum2023inferring}), and asymptotic phases~\cite{wilson2020data,namura2022estimating,yawata2024phase,yamamoto2025gaussian}.
While several studies have proposed methods for estimating the interaction of coupled oscillator systems~\cite{pikovsky2018reconstruction,panaggio2019model,ota2020interaction,matsuki2024network}, these are based on the assumption of pairwise interactions. 
A key open question is whether it is possible to infer the type of interactions (e.g. pairwise, three-body) in real-world systems. 
Recently, several methods for estimating higher-order interactions in complex dynamical systems have been proposed~\cite{casadiego2017model,malizia2024reconstructing,neuhauser2024learning,delabays2025hypergraph}. 
However, identifying the specific type of interaction—whether elements are coupled via pairwise (two-body), three-body, or higher-order group interactions—purely from data remains a significant challenge. 
In particular, determining whether most pairwise couplings are absent (i.e., effectively zero) based on experimental data is a non-trivial task. 
In coupled oscillator systems, both higher order interactions~\cite{skardal2020higher,leon2024higher,marui2025synchronization} and pairwise interactions with additional mechanisms~\cite{acebron2005kuramoto} exhibit rich behaviours such as synchronous states, asynchronous states, and cluster states. 
While the bifurcation structure can differ significantly between pairwise interactions and higher-order interactions, it is not straightforward to determine the interaction type from an experiment. This is because, in experiments, we are unable to fully observe all system variables or to arbitrarily control the bifurcation parameters, and measurements are subject to noise. 
It is therefore essential to develop a method for identifying the type of interaction and inferring the topology and strength of the couplings between the oscillators.

In this study, we address the following question: "Is it possible to identify the interaction type of coupled oscillators from the data?" 
We focus on coupled oscillator systems that include pairwise and three-body interactions. We develop a method based on adaptive LASSO~\cite{zou2006adaptive} for inferring the topology and strength of the individual couplings from time series by assuming the coupling function is known. 
Most existing network reconstruction methods rely on either OLS (Ordinary Least Squares)~\cite{pikovsky2018reconstruction,panaggio2019model,ota2020interaction,malizia2024reconstructing,matsuki2024network} or LASSO (Least Absolute Shrinkage
and Selection Operator)~\cite{han2015robust,shi2021inferring,topal2023reconstructing,delabays2025hypergraph}. In contrast, our proposed method employs the adaptive LASSO, which satisfies desirable oracle properties for large datasets~\cite{zou2006adaptive}.  
We then systematically evaluate the performance of the proposed method using synthetic datasets with different interaction types, coupling probabilities, and observation durations, and compare its performance against two baseline methods: OLS and LASSO. 
Furthermore, we demonstrate the practical applicability of our method by inferring human brain networks~\cite{vskoch2022human}. 
Finally, we present extensions of the method to more general oscillatory systems, including those with limit cycle deformations and those with four-body interactions. 

\section*{Results}

\subsection*{Coupled oscillators with higher order interaction}
We consider a population of $N$ oscillators, which interact via pairwise (two-body) and three-body couplings (i.e. individual interaction)~\cite{leon2019phase,skardal2020higher,nijholt2022emergent,leon2024higher}:
\begin{equation}    	
	\frac{d\phi_i}{d t}=	\omega_i +  f^{(2)}_i \left( \vec{\phi}; {\bf W}^{(2)}_i \right) +  f^{(3)}_i \left( \vec{\phi}; {\bf W}^{(3)}_i \right) + \sigma_i \xi_i(t) ,
		\label{eq:Kuramoto}
\end{equation}   
where $\phi_i \in[0,2 \pi)$ is the phase of the oscillator $i$, $\omega_i$ is its natural frequency, $\sigma^2_i$ is the noise intensity, and $\xi_i(t)$ is the Gaussian white noise, which satisfies ${\rm E}[ \xi_i(t) ] =0$ and ${\rm Cov}[ \xi_i(t) \xi_j(s) ] =\delta_{i, j} \delta(t-s)$, where $\delta_{i, j}$ is Kronecker delta, and $\delta(t-s)$ is the Dirac's delta function. 
It is assumed that the effect of the pairwise and three-body couplings on oscillator $i$ is given by the following equations: 
\begin{eqnarray}		
	f^{(2)}_i \left( \vec{\phi}; {\bf W}^{(2)}_i \right)  &=& 	 \frac{1}{ q^{(2)} N} \sum_{j: j \neq i}^N  W^{(2)}_{i j} \sin \left( \phi_j-\phi_i \right),	 \label{eq:pairwise}	\\
	f^{(3)}_i \left( \vec{\phi}; {\bf W}^{(3)}_i \right)  &=&	 \frac{1}{ q^{(3)} N^2} \sum_{j, l: l>j}^N  W^{(3)}_{i j l}  \{  \sin \left(2 \phi_j-\phi_l-\phi_i \right)
		+\sin \left(2 \phi_l-\phi_i-\phi_j \right)  \},	 \label{eq:threebody} 	
\end{eqnarray}    
where, $\vec{\phi}= (\phi_1, \cdots, \phi_N)$ is the phase of oscillators, and  
${\bf W}^{(2)}_i= (W^{(2)}_{i1}, \cdots,  W^{(2)}_{iN}) $ and ${\bf W}^{(3)}_i= (W^{(3)}_{i12}, \cdots, W^{(3)}_{i1N}, W^{(3)}_{i23}, \cdots, W^{(3)}_{i N-1 N})$ are the vectors of pairwise and three-body coupling strength, respectively. 
The strength of the pairwise coupling from oscillator $j$ to $i$ is denoted by $W^{(2)}_{ij}$, and $W^{(2)}_{ij} = 0$ indicates no coupling between the pair. Similarly, the strength of the three-body coupling from oscillators $j$ and $l$ to $i$ is denoted by $W^{(3)}_{ijl}$, and $W^{(3)}_{ijl} = 0$ indicates no coupling between the triple.  
In this study, we consider an asymmetric random graph (network), that is, every pair or triple of nodes in the network independently couples with a probability of $q^{(2)}$ or $q^{(3)}$, respectively.  
The three-body coupling function~\eqref{eq:Kuramoto} is a symmetric case of the simplicial interaction model~\cite{skardal2020higher}, which is an extension of Kuramoto model~\cite{acebron2005kuramoto}. Similar sinusoidal coupling functions can also be derived from higher-order phase reductions~\cite{ashwin2016hopf,leon2019phase}.

\begin{figure}[h]
    \begin{center}
        \includegraphics[scale=.65]{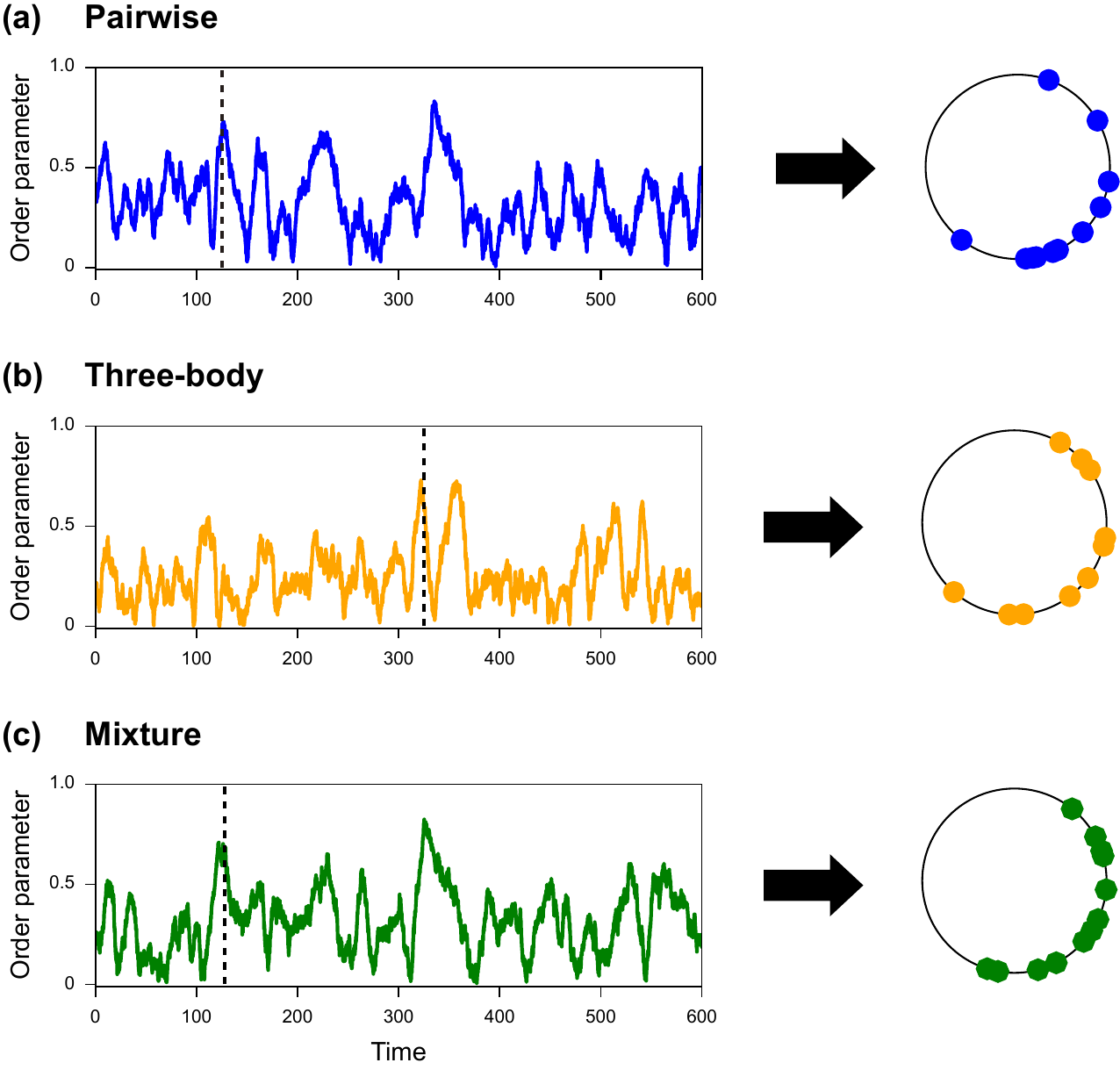}        
        \caption{  \textbf{Synchronization dynamics of weakly coupled oscillators in the presence of white noise (Eq.~\ref{eq:Kuramoto}).}         
        Left: Time series of the Kuramoto order parameter $r$ in a weakly synchronized state for three types of interactions; 
        (a)  Pairwise interaction only,          
        (b)  Three-body interaction only, and  
        (c)  Mixture interactions, where the system has both pairwise and three-body couplings. 
        The average values of the order parameter were $r= 0.33, 0.25$, and $0.32$ for the pairwise, three-body, and mixture interactions, respectively. 
        Right: Snapshots of the phase distribution of the oscillators when they are synchronized at similar levels ($r= 0.67, 0.61$ and $0.67$ for the pairwise, three-body, and mixture interaction, respectively).   
        The weights ($K_2$, $K_3$) and the coupling probabilities ($q^{(2)}$, $q^{(3)}$) were set as follows: 
        A. $K_2= 0.1$, $q^{(2)}= 0.1$and $q^{(3)}= 0.0$, B. $K_3= 0.3$, $q^{(2)}= 0.0$, and $q^{(3)}= 0.1$, and C. $K_2= 0.1$, $K_3= 0.3$, $q^{(2)}= 0.05$ and $q^{(3)}= 0.05$.  
        }   \label{fig:Synchronization}
    \end{center}
\end{figure}

Both pairwise and three-body interactions facilitate synchronization between oscillators. 
Figure~\ref{fig:Synchronization} compares the collective dynamics of the oscillators induced by three types of interactions: 
A. Pairwise interaction, B. Three-body interaction, and C. Mixture of pairwise and three-body interactions. 
For simplicity, the interaction is described as the random graph with homogeneous coupling strength: 
A. for the pairwise interaction, $W^{(2)}_{ij}= K_2$ with a probability $q^{(2)}$, otherwise $W^{(2)}_{ij}= 0$, 
B. for the three-body interaction, $W^{(3)}_{ijl}= K_3$ with a probability $q^{(3)}$, otherwise $W^{(3)}_{ij}= 0$, 
and C. for the mixture interaction, $W^{(2)}_{ij}= K_2$ with a probability $q^{(2)}$, otherwise $W^{(2)}_{ij}= 0$, and $W^{(3)}_{ijl}= K_3$ with a probability $q^{(3)}$, otherwise $W^{(3)}_{ij}= 0$. 

\clearpage

In order to quantify the level of synchrony of the system, we calculate the Kuramoto order parameter $r:= \left| \sum_{k=1}^N e^{\sqrt{-1} \theta_k} / N \right|$, where $|z|$ is the modulus of a complex number $z$. 
The order parameter reaches its maximum value of $r= 1$ when all the oscillators in a system are synchronized in phase, while it takes its minimum value of $r= 0$ when the phases of the oscillators are uniformly distributed. 
We calculate the time series of the order parameter, $r(t)$, for the three interaction types, and find that the time series exhibit similar behaviour among the oscillator populations with three interaction types (Fig.~\ref{fig:Synchronization}, left). 
We then compare the phase distributions of the oscillators at different times with similar levels of synchrony (Fig.~\ref{fig:Synchronization}, right). The snapshots of the phase distribution were found to be similar for the three types of interactions. 
These results imply that it is challenging to identify the interaction type only from the order parameter $r(t)$ when the oscillators are noisy. 
Here we address the following two questions: 
\begin{itemize}
	\item		Can we identify the type of interaction (e.g., pairwise, three-body, and mixture) among the oscillators based on the phase time series $\{ \phi_i(t) \}$ ($i= 1, 2, \cdots, N$) of oscillators? 
	\item		How accurately can we infer the topology and the weight of the couplings (i.e. the individual interaction) from the phase time series?
\end{itemize}

\subsection*{Adaptive LASSO method for inferring the interaction network from time series}
Here, we aim to identify the type of interaction between the oscillators and infer the topology and strength of the couplings (i.e. individual interaction) from the phase time series $\{ \phi_i(t) \}$ of oscillator $i$ ($i= 1, 2, \cdots , N$). 
Suppose that the phase time series is obtained at $L$ points at regular intervals $h$: $\phi_i(h), \phi_i(2h), ..., \phi_i(Lh)$. 
The derivative $d\phi_i/dt$ at time $t$ is approximated by the difference,  $\frac{\Delta \phi_i}{\Delta t}(t)=: (\phi_i(t+h)- \phi_i(t))/h$.  

In this paper, we develop a method based on the adaptive LASSO~\cite{zou2006adaptive} for inferring the couplings from time series, assuming the functional form of the coupling function (\ref{eq:pairwise},\ref{eq:threebody}) is known. 
The model parameters, including the natural frequency ($\omega_i$) and the coupling parameters (${\bf W}^{(2)}_i$ and ${\bf W^{(3)}_i}$), are estimated by fitting the dynamical model~\eqref{eq:Kuramoto} for an oscillator $i$. This estimation procedure is repeated for all oscillators ($i= 1, 2, \cdots , N$). 
First, a preliminary estimate of the parameters ($\omega_i$, ${\bf W}^{(2)}_i$, and ${\bf W}^{(3)}_i$) is obtained by the Ordinary Least Squares (OLS) method~\cite{malizia2024reconstructing}, which minimizes the squared error: 
\begin{equation}
    E_{\rm OLS} =		\sum_{k=0}^{L-1}	\left(	  \frac{\Delta \phi_i}{\Delta t}(kh)	 - \omega_i -  
	f^{(2)}_i( {\bf \phi}(kh); {\bf W}^{(2)}_i ) -	f^{(3)}_i( {\bf \phi}(kh); {\bf W}^{(3)}_i )	\right)^2,  			
	\label{eq:LS}
\end{equation}
where $h$ is the sampling interval. The minimization is performed by finding the local minimum at which the derivative is zero. 

Next, the estimate is refined through minimizing the cost function: 
\begin{equation}
    E_{\rm AL} =  E_{\rm OLS}+ \alpha	    
    \left\{    \sum_{j: j \neq i}	     \left|  c^{(2)}_{ij}  W^{(2)}_{ij}  \right| + 
    \sum_{j, l: l >j, j \neq i, l  \neq i}     \left|  c^{(3)}_{ijl} W^{(3)}_{ijl}   \right| \right\},   	
	\label{eq:A-LAS}
\end{equation}
where $|x|$ represents the absolute value of a real number $x$. 
The hyperparameters $c^{(2)}_{ij}$ and $c^{(3)}_{ijl}$ are defined as follows: 
$c^{(2)}_{ij}= 1/ \tilde{W}^{(2)}_{ij}$ and $c^{(3)}_{ijl}= 1/ \tilde{W}^{(3)}_{ijl}$, where $\tilde{W}^{(2)}_{ij}$ and $\tilde{W}^{(3)}_{ijl}$ are the preliminary estimates obtained by minimizing the squared error $E_{\rm OLS}$. 
For a fixed regularization parameter $\alpha$, the refined parameters ($W^{(2)}_{ij}$ and $W^{(3)}_{ijl}$) are estimated using Least Angle Regression (LARS)~\cite{efron2004least}.   
The optimal value of $\alpha$ is selected by minimizing the Bayesian Information Criterion (BIC), which is implemented in the \textit{LassoLarsIC} function from \textit{scikit-learn} \cite{JMLR:v12:pedregosa11a}. 
This adaptive LASSO method is capable of both identifying the presence of  coupling (e.g, whether $W^{(2)}_{ij}= 0$ or not) and estimating the coupling strengths.

\begin{figure}[h]
\begin{center}
    \includegraphics[scale=.7]{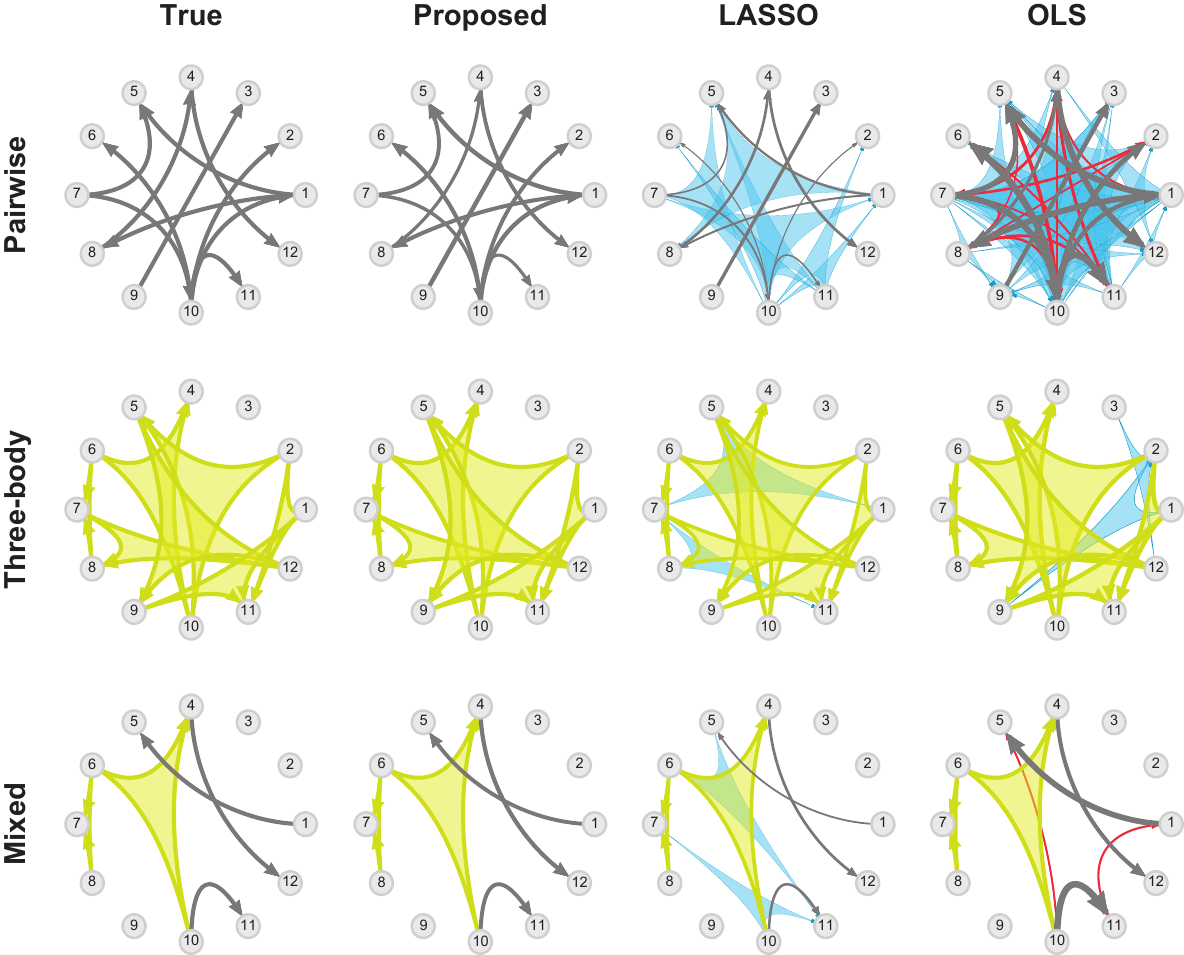}
    \vspace{1cm}     
    \caption{	\textbf{Inferring the pairwise and three-body couplings from the phase time series.}   
   The couplings (i.e. individual interaction) inferred from synthetic data sets using the proposed method, LASSO, and OLS methods are compared with the true couplings. 
   Three synthetic datasets are analysed: pairwise interaction (Top), three-body interaction (Middle), and mixture interaction (Bottom). 
   The coupling probabilities were set to $q^{(2)}= 0.06$, $q^{(3)}= 0.00$ in the pairwise interaction, $q^{(2)}= 0.00$, $q^{(3)}= 0.01$ in the three-body interaction, and $q^{(2)}= 0.03$, $q^{(3)}= 0.005$ in the mixture interaction. 
   Correctly inferred (true positive) pairwise and three-body couplings are shown in black and yellow, respectively, while incorrectly inferred (false positive) pairwise and three-body couplings are shown in red and blue, respectively.     
    } 
     \label{fig:Inference_result}
\end{center}
\end{figure}

Figure~\ref{fig:Inference_result} compares the inferred couplings from the proposed method with those from two baseline methods, LASSO (Least Absolute Shrinkage and Selection Operator) and OLS (Ordinary Least Squares) with a statistical test (see Method for details). 
LASSO is a popular method for network inference problems in general~\cite{han2015robust}(and others). It has also been applied to inference problems of the same type as ours, namely network inference from time-series~\cite{delabays2025hypergraph}. 
Extensions of LASSO, such as Signal LASSO~\cite{shi2021inferring}, have been successfully applied to network inference problems. However, to our knowledge, the adaptive LASSO has not been used before. 
The proposed method was able to infer the topology of the pairwise and three-body couplings with complete accuracy (Fig.~\ref{fig:Inference_result}, Proposed), with no false positives and no false negatives.  
OLS produced a considerable number of false positives for both pairwise and three-body couplings (Fig.~\ref{fig:Inference_result}, OLS). When the true interaction was pairwise only, the number of false positives was 16 (12 \%)  and 23 (3.5 \%) for pairwise and three-body couplings, respectively. 
While LASSO gave a lower number of false positives than OLS, it gave the false positives when the true interaction was pairwise (Fig.~\ref{fig:Inference_result}, LASSO). 
When the true interaction was pairwise only, the number of false positives was 0 (0 \%)  and 12 (1.8 \%) for pairwise and three-body couplings, respectively. 
Furthermore, the width of the arrow indicates that LASSO underestimates the coupling strength. 
These results suggest that the proposed method performs better than OLS and LASSO in inferring the couplings from the time series. In the next subsection, we systematically compare the inference performance of the proposed method with that of the baseline methods. 

\subsection*{Evaluation of the proposed method}

We evaluate the performance of the proposed method for identifying the couplings (i.e. individual interaction) between the oscillators from synthetic phase time series.    
The synthetic data are generated by simulating a population of oscillators~\eqref{eq:Kuramoto} using the Euler-Maruyama method with a time step of $\delta t= 0.02$. 
Regarding the interaction, we consider three types of random graphs with constant strength: pairwise, three-body, and mixture interactions, as shown in Fig.~\ref{fig:Synchronization}. 
In this subsection, we assume that the probabilities of the existence of pairwise and three-body couplings are equal, $q:= q^{(2)}= q^{(3)}$, in the mixture interaction. The coupling strength is either zero or a constant value, $K_2$ for pairwise couplings and $K_3$ for three-body couplings, respectively. 

Firstly, we test whether it is possible to distinguish the type of interaction (pairwise, three-body, or mixture) from the time series. To do this, we obtain the probabilities of the existence of pairwise and three-body couplings, $\hat{q}^{(2)}$ and $\hat{q}^{(3)}$, by calculating the proportion of non-zero values in the inferred coupling parameters: $\hat{W}^{(2)}_{ij}$ and $\hat{W}^{(3)}_{ijl}$.  
The proposed method, as well as three baseline methods (naive OLS, OLS with statistical test, and LASSO: see Methods for details) are then used to infer the coupling parameters from time series. 
The null hypothesis that the pairwise and three-body coupling probabilities are equal: $q^{(2)}= q^{(3)}$ is then tested using the Z-test (see Methods for details). 
Rejection of the null hypothesis indicates that the pairwise and three-body coupling probabilities are significantly different. 
Therefore, if the null hypothesis is rejected, we conclude that the interaction with the higher coupling probability is dominant. In other words, if $\hat{q}^{(2)} > \hat{q}^{(3)}$, we conclude that the interaction is pairwise dominant. Conversely, if $\hat{q}^{(2)} < \hat{q}^{(3)}$, we conclude that the interaction is three-body dominant. Additionally, if the null hypothesis is not rejected, we conclude that the interaction is a mixture.

Table~\ref{tab:Interact_Type} shows the accuracy of inferring the interaction type from the time series, which were generated from 20 populations of each pairwise, three-body, and mixture interaction. All the methods achieve an accuracy of over 80 \% in identifying the interaction type. 
In particular, the proposed method and LASSO achieve perfect identification of the interaction type, i.e. 100 \% accuracy.
Table~\ref{tab:Interact_Prob} shows the inferred probabilities of  the existence of pairwise and three-body couplings ($\hat{q}^{(2)}$ and $\hat{q}^{(3)}$) obtained from the same data as in Table~\ref{tab:Interact_Type}. 
In the case of synthetic data with pairwise interaction, the proposed method demonstrates the highest accuracy in inferring the coupling probability. In contrast, LASSO underestimated and OLS overestimated the probability of pairwise coupling. 
In addition, both LASSO and OLS exhibited an overestimation of the probability of three-body coupling by 0.6-0.7 \%, and the naive OLS exhibited an overestimation of more than 9.0 \%. 
In the case of synthetic data with three-body interaction, LASSO was the most accurate in inferring the probability of the pairwise couplings, while the proposed method was the most accurate in inferring that of the three-body couplings.
In contrast, Naive OLS overestimated the probability of pairwise (5.1 \%) and three-body (3.7 \%) couplings, and OLS underestimated the probability of the three-body couplings. 
In the case of synthetic data with mixture interaction, LASSO achieved the most accurate inference for the probability of the pairwise couplings, while the proposed method achieved the most accurate inference for that of the three-body couplings. 
OLS overestimates the probability of the pairwise couplings, and it underestimates that of the three-body couplings. Naive OLS  overestimated the probability of pairwise (7.3 \%) and three-body (5.3 \%) couplings. 
Overall, the naive OLS method cannot infer the coupling probabilities in the present situation.
In contrast, a recent study~\cite{malizia2024reconstructing} has shown that OLS method can infer the higher-order interactions from time series accurately. 
The discrepancy between these findings can be attributed to the difference in the situation being considered. 
Previous research has focused on scenarios without noise, while this study considers a system of noisy coupled oscillators.
Tables~\ref{tab:Interact_Prob} indicates that sophisticated methods, including OLS with a statistical test and LASSO~\cite{tibshirani1996regression} (see Method for details), 
are essential for distinguishing between weak couplings and no couplings in noisy situations. In the following, the proposed method is compared with two sophisticated methods, i.e. OLS with statistical test and LASSO.  

\begin{figure}[t]
\begin{center}
    \includegraphics[scale=.7]{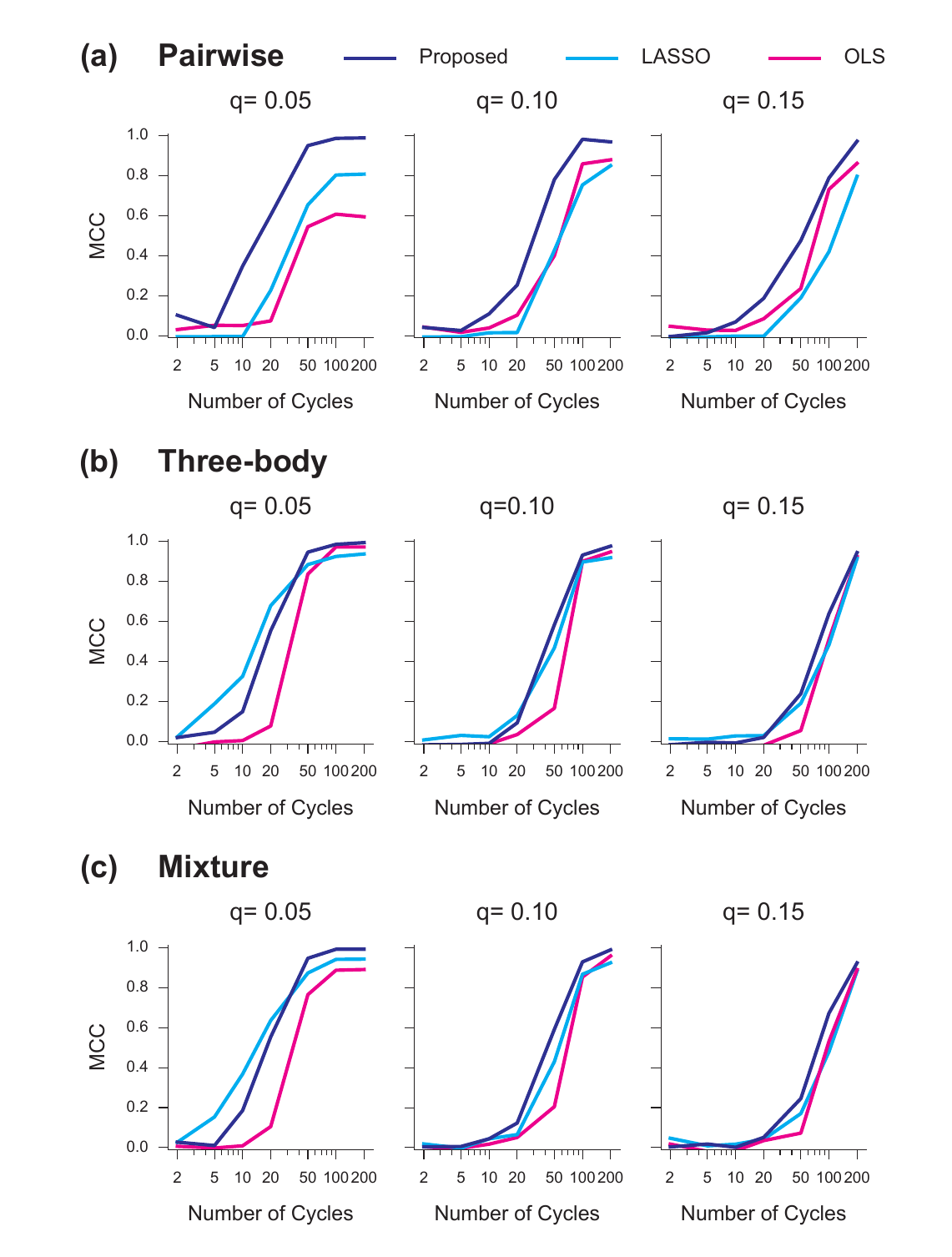}
    \caption{	\textbf{Effect of observation time on inference accuracy.} 
    Performance of coupling inference (MCC) was calculated for the three types of interaction: (a) Pairwise ($q^{(2)}= q$, $q^{(3)}= 0$), (b) Three-body ($q^{(2)}= 0$, $q^{(3)}= q$), and (c) Mixture of pairwise and three-body interactions ($q^{(2)}= q/2$, $q^{(3)}= q/2$). The parameter of the coupling probability $q$ was set to 0.05, 0.10 and 0.15, on the left, centre and right panel, respectively. 
    }   \label{fig:MCC_vs_Cycle}
\end{center}
\end{figure}

We evaluate the performance in inferring whether or not there was a coupling (i.e. individual interaction) in each pair or triple by calculating the Matthews Correlation Coefficient (MCC)~\cite{Kobayashi2019}, which is defined as follows
\begin{eqnarray}	
  	MCC=	\frac{N_{\rm TP} N_{\rm TN}- N_{\rm FP} N_{\rm FN}  }{ \sqrt{(N_{\rm TP}+ N_{\rm FP}) (N_{\rm TP}+ N_{\rm FN}) (N_{\rm TN}+ N_{\rm FP}) (N_{\rm TN}+ N_{\rm FN})}  },	
\end{eqnarray}	
where $N_{\rm TP}$, $N_{\rm TN}$, $N_{\rm FP}$, and $N_{\rm FN}$ are the numbers of true positives (TPs), true negatives (TNs), false positives (FPs), and false negatives (FNs) of the pairwise and three-body couplings, respectively.   
MCC values range from -1 to 1. If all the couplings are correctly inferred, the MCC value is 1. If the couplings are randomly inferred with the true coupling probability, the expected value of MCC is 0.
%
Figure~\ref{fig:MCC_vs_Cycle} shows the dependence of the MCC on the observation duration for the three interaction types (pairwise, three-body, and mixture) with three coupling probabilities: $q= 0.05, 0.10$, and $0.15$. 
As the coupling probability increases, a longer observation period is required for an accurate inference. 
For example, 50 cycles are sufficient to accurately ($MCC > 0.8$) infer the couplings for the synthetic data of a sparse interaction ($p=0.05$). In contrast, more than 100 cycles are necessary to infer the couplings for the synthetic data of a dense interaction ($p=0.15$). 
Overall, when the observation period is sufficiently long, the proposed method demonstrates superior performance in terms of accuracy compared to LASSO and OLS. 
Furthermore, the proposed method outperforms LASSO and OLS in inferring the couplings across a broad range of coupling strengths, $K_2$ and $K_3$ (see Supplementary Note 1). 

\clearpage

\begin{figure}[t]
    \begin{center}
        \includegraphics[scale=.7]{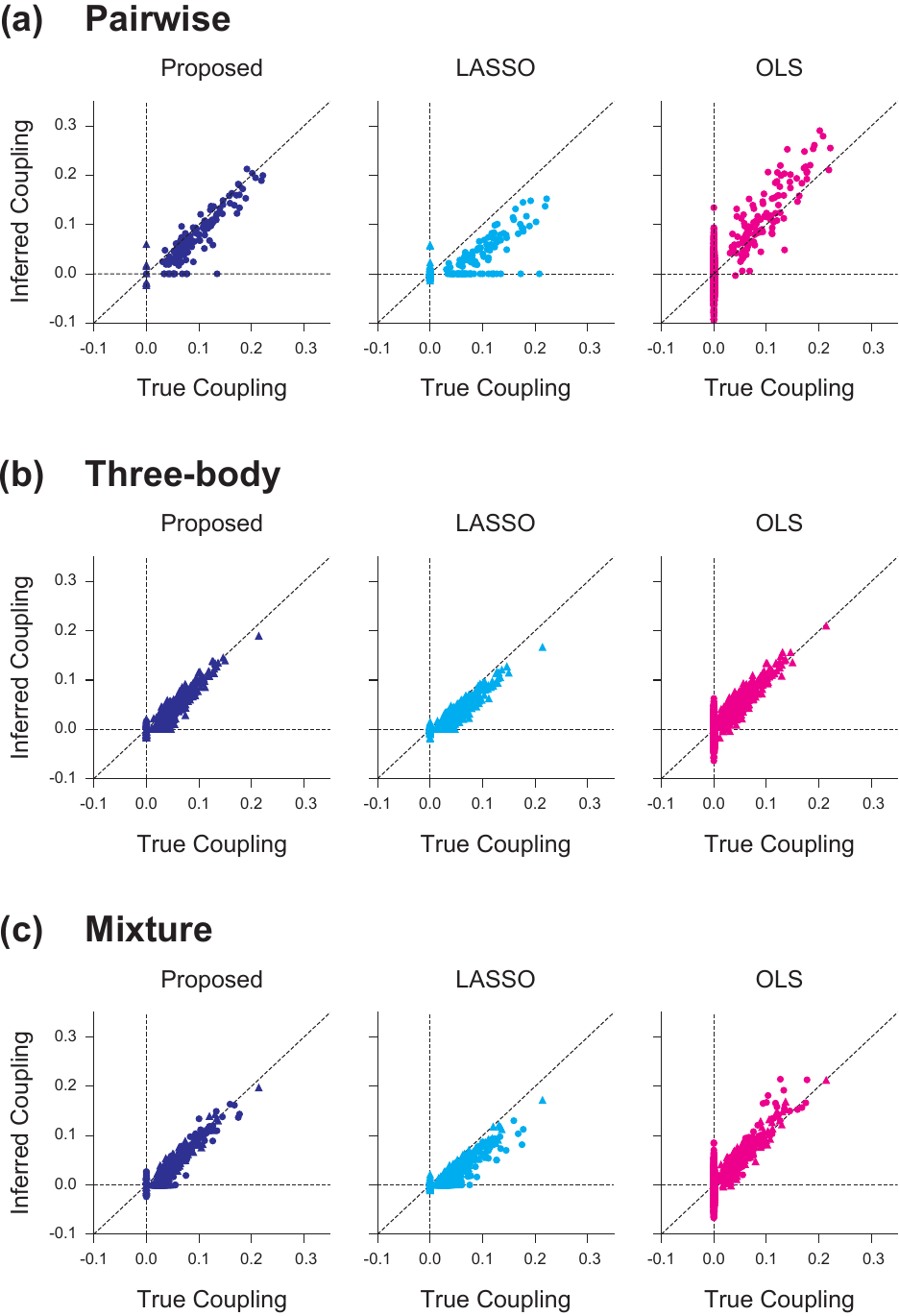}
        \caption{	\textbf{Inferring coupling strength from time series.}   
        Inferred coupling strengths ($\hat{W}^{(2)}_{ij}$ and $\hat{W}^{(3)}_{ijl}$) obtained by the three methods (Left: proposed method, Centre: LASSO, and Right: OLS) are compared with the true strengths ($W^{(2)}_{ij}$ and $W^{(3)}_{ijl}$) for the three types of interaction: (a) Pairwise ($q^{(2)}= 0.1$, $q^{(3)}= 0$), (b) Three-body ($q^{(2)}= 0$, $q^{(3)}= 0.1$), and (c) Mixture of pairwise and three-body interactions ($q^{(2)}= 0.05$, $q^{(3)}= 0.05$).          
        The points on the first quadrant represent true positives, while those on the non-zero $x$-axis and $y$-axis represent false negatives and false positives, respectively. 
        The points above (below) the diagonal line indicate an overestimation (underestimation) of the coupling strength. The circle and triangle points represent the inferred results for pairwise and three-body interactions, respectively. 
        }   \label{fig:Est_Coupling}	
    \end{center}	
\end{figure}

Finally, we evaluate the inference performance obtained from the coupled oscillators with highly heterogeneous coupling strength. Such interactions are observed in neural circuits in the brain~\cite{buzsaki2014log,Kobayashi2019}. 
We simulate a population of oscillators~\eqref{eq:Kuramoto} whose coupling strength is given by the Erd{\" o}s-R{\' e}nyi random graph with a log-normal distribution of the strength.  
Figure~\ref{fig:Est_Coupling} shows the inference result of the coupling strength by the three methods (our proposed method, LASSO, and OLS). 
True negatives are plotted at the origin, 
false negatives are plotted on the positive part of the $x$-axis, 
false positives are plotted on the positive part of the $y$-axis, 
and true positives are plotted in the first quadrant.  
Again, the performance of coupling inference is evaluated using MCC (Table~\ref{tab:MCC_small_LN}). 
The MCC values of the proposed method were higher than those of LASSO and OLS, indicating that the proposed method outperforms LASSO and OLS in terms of coupling inference. 
Furthermore, we examine the performance in inferring the coupling strengths.  
The coupling strengths inferred by OLS are distributed above the diagonal line, indicating that OLS overestimates the strength. Conversely, the coupling strengths inferred by LASSO are distributed below the diagonal line, indicating that LASSO underestimates the strength. 
The performance in inferring the strength is evaluated by calculating the mean squared error: $\epsilon^2= \frac{1}{M} \sum_{C} (w_C- \hat{w}_C )^2$, where $C= \{ i, j \}$ or $\{ i, j, l \}$ represents the index of  pairwise or three-body couplings, $w_C$ is a true coupling strength, i.e. $W^{(2)}_{ij}$ or $W^{(3)}_{ijl}$, $\hat{w}_C$ is its inferred value, and $M= N(N-1)$ or $N (N-1) (N-2)/2$ is the maximal number of pairwise or three-body couplings, respectively.  
The mean squared error obtained by the proposed method was lower than that obtained by LASSO and OLS (Table \ref{tab:MSE_small_LN}). 
As demonstrated in Figure~\ref{fig:Est_Coupling}, OLS tends to overfit the coupling strength for uncoupled pairs or triples, which is a possible reason for the large error. 
This result indicates that the proposed method outperforms LASSO and OLS in inferring the coupling strength.

\begin{figure}[t]
\begin{center}
    \includegraphics[scale=.65]{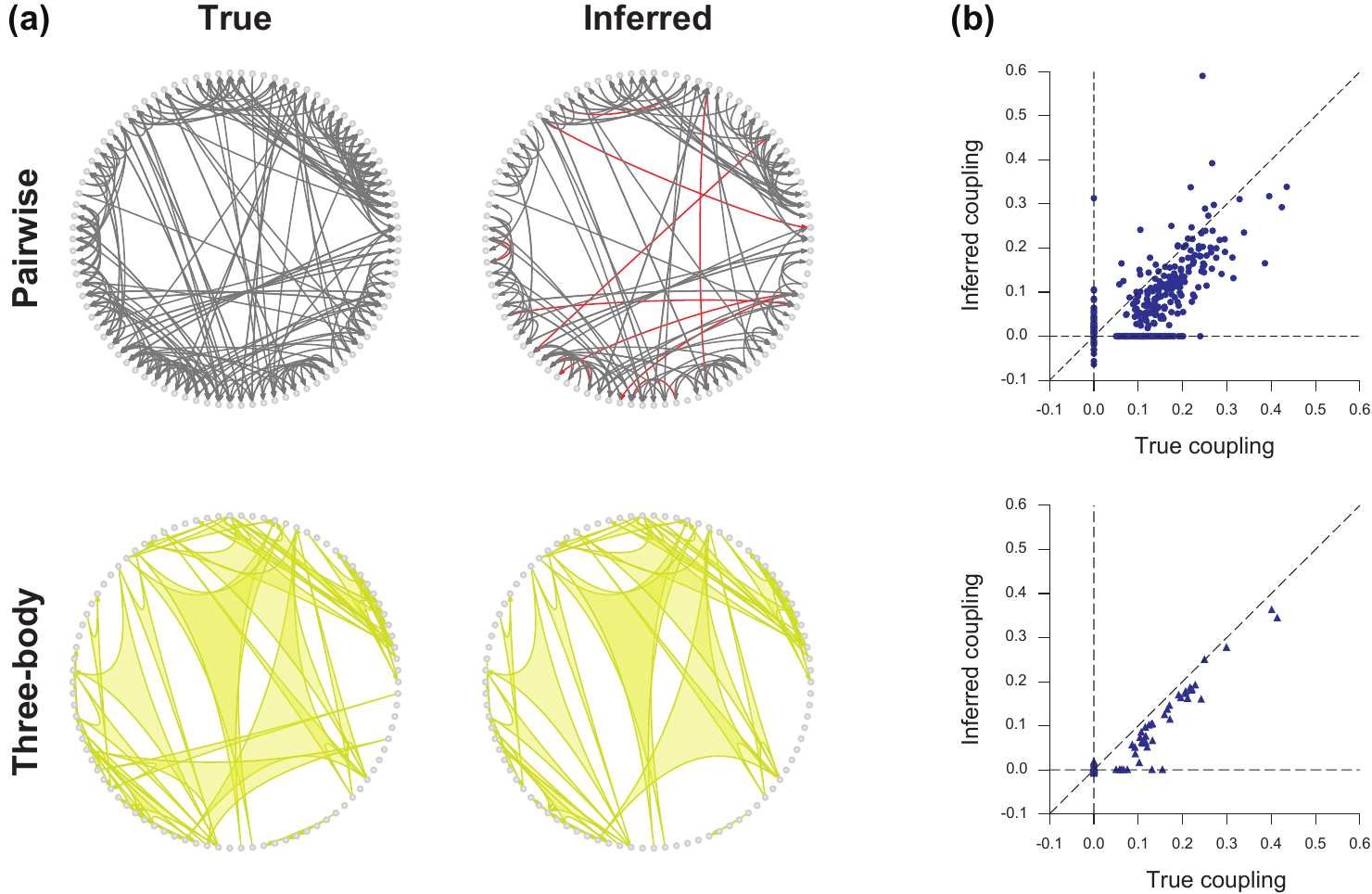}
    \caption{ 
    \textbf{Application to a human brain network (Subject ID: S038).}  
    (a) Inferred coupling network. 
    The proposed method was applied to infer the couplings of the human brain network from time series data. The true network (left) is compared with the inferred network (right). 
    Top panels display pairwise couplings; bottom panels show three-body couplings. 
    Only strong couplings with magnitudes greater than 0.05 are visualized. 
    Correctly inferred (true positive) couplings are shown in black (top: pairwise) and yellow (bottom: three-body). 
    Incorrectly inferred (false positive) couplings are shown in red (top: pairwise) and blue (bottom: three-body).         
    (b) Inferred coupling strengths.    
    Scatter plots show inferred versus true coupling strengths for pairwise (top) and three-body (bottom) interactions. 
    Points on the non-zero x-axis represent false negatives (missed true couplings), while those on the non-zero y-axis represent false positives (spurious couplings). 
    }   \label{fig:Large_Pop}
\end{center}
\end{figure}

\subsection*{Application to a real-world network} 

As demonstrated in previous subsections, the proposed method accurately infers both the interaction type and the couplings from time series using synthetic data sets with a small number of oscillators (Table \ref{tab:Interact_Type}, Figs. \ref{fig:Inference_result} and \ref{fig:MCC_vs_Cycle}: 12 oscillators). 
To further assess the feasibility of our method, we apply it to reconstruct the human brain network. 
We use the human brain structural connectivity dataset~\cite{vskoch2022human} obtained from healthy subjects. 
The network nodes correspond to 90 cortical regions of interest (ROIs), as defined by the Automatic Anatomical Labeling (AAL) atlas. 
Structural connectivity matrices are derived from probabilistic tractography applied to diffusion MRI data. 
The human brain networks are constructed as follows: the link from the node $i$ to $j$ is connected when the structural connectivity  from $i$ to $j$ is at least $0.05$. This network is directed because the structural connectivity matrices were asymmetric. 
Pairwise coupling strengths $W^{(2)}_{ij}$ are assigned to the links using a truncated lognormal distribution, which mimics the distribution observed in neuronal circuits~\cite{buzsaki2014log,Kobayashi2019}. All values are ensured to be greater than $0.05$.  
Three-body couplings are assigned to triplets of nodes where pairwise links exist between each pair (i.e., between nodes $i$ and $j$, $j$ and $l$, and $l$ and $i$). The direction of each three-body coupling is determined randomly with equal probability. Here, only the three-body couplings were added, while the pairwise couplings remained unchanged.
Three-body coupling strengths $W^{(3)}_{ijl}$ are also drawn from a truncated lognormal distribution, with all values exceeding $0.05$.  
Finally, phase time series are generated by simulating the coupled oscillators~(\ref{eq:Kuramoto}) using the pairwise and three-body coupling strengths derived from the human brain data. The inference is performed using 200 cycles of simulated time series. 

Figure~\ref{fig:Large_Pop}A shows the inference results for strong couplings of a human brain network (Subject ID: S038). 
While our method detects some false positives in pairwise interactions even for strong couplings, it does not detect any false positives in three-body interactions. 
The overall MCC was 0.694 with 90 false positives and 102 false negatives.  
Furthermore, our method accurately estimates the coupling strengths (Fig.~\ref{fig:Large_Pop}B). The mean squared error $\epsilon^2$ of the inferred coupling strengths across all couplings was $1.15 \times 10^{-6}$.  
Table~\ref{tab:MCC_HB_LN} compares the inference performance on the human brain networks (four subjects) across three methods: our proposed method, LASSO, and OLS. Our method consistently outperforms the approaches in reconstructing the coupling structure. These results highlight the potential of our method for inferring real-world network structures from time series. 
\clearpage

\subsection*{Extensions of the proposed method}

We have examined the coupling inference problem for a simple oscillator system: Kuramoto-type coupled oscillators with both pairwise and three-body interactions. 
While the Kuramoto-type model can qualitatively explain various synchronous phenomena~\cite{galan2005efficient,acebron2005kuramoto}, the coupling function of weakly coupled oscillators can be described as a sum of basis functions~\cite{kuramoto1984chemical,nakao2016}. 
In this subsection, we further extend the proposed method to more general oscillator systems.

We begin with a system of oscillators exhibiting phase-dependent amplitude~\cite{bick2024higher}, capable of describing a wide range of real-world oscillatory systems.  
Assuming symmetric three-body couplings, a weakly coupled oscillator system can be approximately described by the following equation (see Supplementary Note 2 for details): 
\begin{equation}    	
    \frac{d\phi_i}{d t}=	\omega_i +  f^{(2)}_i \left( \vec{\phi}; {\bf W}^{(2)}_i \right) +  f^{(3, 1)}_i \left( \vec{\phi}; {\bf W}^{(3, 1)}_i \right) + f^{(3, 2)}_i \left( \vec{\phi}; {\bf W}^{(3, 2)}_i \right) + \sigma_i \xi_i(t) ,
    \label{eq:Osc_gen_3-body}
\end{equation} 
where $\vec{\phi}= (\phi_1, \cdots, \phi_N)$ is the vector of oscillator phases and 
${\bf W}^{(2)}_i= (W^{(2)}_{i1}, \cdots,  W^{(2)}_{iN}) $, 
${\bf W}^{(3, 1)}_i= (W^{(3, 1)}_{i12}, \cdots, W^{(3, 1)}_{i1N}, W^{(3, 1)}_{i23}, \cdots,$ $W^{(3, 1)}_{i N-1 N})$, and 
${\bf W}^{(3, 2)}_i= (W^{(3, 2)}_{i12}, \cdots, W^{(3, 2)}_{i1N}, W^{(3, 2)}_{i23}, \cdots, W^{(3, 2)}_{i N-1 N})$
are the vectors of pairwise and three-body coupling strength, respectively. 
Using the Brent method~\cite{matsuki2024network}, the inference problem 
for coupled oscillators with phase-dependent amplitude can be reduced to the estimation problem described by Eq. (\ref{eq:Osc_gen_3-body}) (see Supplementary Note 2). 
To evaluate performance, we apply the proposed framework to a small oscillator system ($N=12$) with coupling probabilities: $q^{(2)} = 6.8\%$, $q^{(3,1)} = 2.0\%$, and $q^{(3,2)} = 1.1\%$. 
The resulting MCC of $0.899$ -- substantially higher than those obtained by baseline methods (LASSO: 0.807; OLS: 0.702) -- demonstrates the effectiveness of our framework in inferring coupling structures in a broader class of oscillatory systems.

Next, we investigate whether the proposed framework can infer higher-order interactions beyond triadic couplings. 
Specifically, we consider a system that incorporates four-body interactions (see Supplementary Note 3 for details): 
\begin{equation}    	
    \frac{d\phi_i}{d t}=	\omega_i +  f^{(2)}_i \left( \vec{\phi}; {\bf W}^{(2)}_i \right) +  f^{(3)}_i \left( \vec{\phi}; {\bf W}^{(3)}_i \right) +  f^{(4)}_i \left( \vec{\phi}; {\bf W}^{(4)}_i \right) + \sigma_i \xi_i(t) ,
    \label{eq:Kuramoto_4-body}
\end{equation} 
where $\vec{\phi}= (\phi_1, \cdots, \phi_N)$ is the vector of oscillator phases and 
${\bf W}^{(2)}_i= (W^{(2)}_{i1}, \cdots,  W^{(2)}_{iN}) $, ${\bf W}^{(3)}_i= (W^{(3)}_{i12}, \cdots, W^{(3)}_{i1N}, W^{(3)}_{i23}, \cdots, W^{(3)}_{i N-1 N})$, and 
${\bf W}^{(4)}_i= (W^{(4)}_{i123}, \cdots, W^{(4)}_{i12N}, W^{(4)}_{i134}, \cdots, W^{(4)}_{i N-2 N-1 N})$
are the vectors of pairwise, three-body, and four-body coupling strength, respectively. 
As with the three-body case, our method can be extended to handle four-body interactions (see Supplementary Note 3). 
To assess performance, we apply the method to a system of $N=12$ oscillators with coupling probabilities $q^{(2)} = 5.3\%$, $q^{(3)} = 1.4\%$, and $q^{(4)} = 0.5\%$. 
The resulting MCC of $0.782$ outperforms the baseline methods (LASSO: $0.577$; OLS: $0.307$). 
This result demonstrates that our framework is capable of inferring four-body interactions. However, the inference performance for four-body couplings is lower than that for pairwise and three-body interactions, suggesting that higher-order inference is more challenging and may require longer time series.

\section*{Discussion}

In this study, we have addressed the question of whether it is possible to distinguish between the types of interaction (i.e. pairwise, three-body, and mixture interactions, all of which contribute to synchronisation) from time series data. 
As seen in Figure~\ref{fig:Synchronization}, the dynamics of the systems with the three types of interaction were similar, indicating the need for advanced methods to accurately distinguish between them.    
In order to identify the interaction type, we have developed a method based on the adaptive LASSO for inferring the couplings (i.e. individual interaction) from phase time series. The proposed method is capable of inferring the interaction type and the probability of pairwise and three-body couplings from time series
(Tables~\ref{tab:Interact_Type} and~\ref{tab:Interact_Prob}). 
Our method has been shown to outperform two baseline methods (LASSO and OLS) in inferring the topology (Figs.~\ref{fig:Inference_result} and \ref{fig:MCC_vs_Cycle}, and Table~\ref{tab:MCC_small_LN}) and strength (Fig.~\ref{fig:Est_Coupling} and Table~\ref{tab:MSE_small_LN}) of couplings. 
Furthermore, we have demonstrated the practical utility of the proposed method through its application to human brain networks (Fig.~\ref{fig:Large_Pop} and Table~\ref{tab:MCC_HB_LN}). 
Finally, we have demonstrated that the proposed method can be extended to more general oscillatory systems, including those exhibiting limit cycle deformations and systems with four-body interactions.

Our study is based on three main assumptions. 
Firstly, we assume that phase data $\{ \phi_i(t) \}$ ($i= 1, 2, ..., N$) is available. 
It should be noted that in many real-world cases, an oscillatory signal is observed instead of the phase; however, the phase can still be reconstructed from the observed signal. 
In cases where the observed signal is similar to a sinusoidal wave, the Hilbert transform can be used to reconstruct the phase~\cite{pikovsky2003synchronization,kralemann2008phase,gengel2019phase,matsuki2023extended}. In cases where marked events such as neural spikes are observed, linear interpolation is often used to reconstruct the phase~\cite{galan2005efficient,tsubo2007,nakae2010bayesian}. 
Second, we assume that the functional forms of the coupling functions are known: $f^{(2)}_i \left( \vec{\phi}; {\bf W}^{(2)}_i \right)$ and $f^{(3)}_i \left( \vec{\phi}; {\bf W}^{(3)}_i \right) $ (Eqs. \ref{eq:pairwise} and \ref{eq:threebody}). 
These coupling functions can be derived by applying the phase reduction technique to weakly coupled Stuart-Landau oscillators~\cite{leon2019phase}, thereby supporting the validity of the assumption in the vicinity of the Hopf bifurcation. 
When the oscillators are strongly synchronised, 
the model with first-order harmonic couplings is capable of reproducing small fluctuations around the synchronised state~\cite{matsuki2024network}. 
However, if the oscillators are not synchronised, coupling functions with higher harmonics would be suitable to describe a global coupling function~\cite{pikovsky2018reconstruction,panaggio2019model,ota2020interaction}. 
Thirdly, we have focused on the pairwise and three-body interactions. 
It is straightforward to extend the present method to systems with four- or higher-order interactions. However, the number of parameters is $O(N^k)$ and the required computational cost is $O(N^{3k})$, which depends exponentially on the order $k$ of the interaction. In the case of 90 oscillators, we estimated 8,000 parameters for pairwise interactions and 352,000 parameters for three-body interactions (Fig.~\ref{fig:Large_Pop}).   
From a computational perspective, it would be challenging to infer four-body or higher-order interactions from data for a large population (100 or more) of oscillators. Consequently, it is an important future study to develop more efficient algorithms for inferring higher-order interactions. 
Another direction for future research is the development of an efficient inference algorithm that focuses on global measures of the interaction network, such as the eigenvalues and eigenvectors of its Laplacian matrix.

In this study, we have demonstrated that it is possible to distinguish between pairwise and three-body interactions and to accurately infer the topology and strength of couplings from time series data, even in the presence of strong noise. Notably, the inferred interactions can characterize causal relationships between oscillators. 
One promising avenue for future research would be to apply the proposed method to physiological data from cardiorespiratory or neuronal systems. The inferred interactions could serve as potential disease biomarkers~\cite{herzog2022genuine}. 
While the current study focuses on the weakly synchronized regime, extending the method to accommodate highly synchronous systems~\cite{matsuki2024network} remains an important challenge for future work.

%
%
\section*{Methods}

\subsection*{Coupled oscillators with higher-order interaction}

We simulate a population of $N$ oscillators that interact through pairwise and/or three-body couplings (see Eqs. \ref{eq:Kuramoto}, \ref{eq:pairwise} and, \ref{eq:threebody}) for details). 
First, we consider a population with homogeneous coupling strength, i.e., $K_{ij}^{(2)}= \bar{K}_2$ and $K_{ijl}^{(3)}= \bar{K}_3$ (Figs.~\ref{fig:Synchronization}, \ref{fig:Inference_result}, and \ref{fig:MCC_vs_Cycle}, and Tables~\ref{tab:Interact_Type} and ~\ref{tab:Interact_Prob}). 
Then, we consider oscillators with heterogeneous coupling strength (Figs. \ref{fig:Est_Coupling} and \ref{fig:Large_Pop}), each component of the weight matrix or tensor is drawn from a log-normal distribution inspired by the couplings measured from neuronal circuits~\cite{buzsaki2014log,Kobayashi2019}: 
\begin{eqnarray}
    	P_{LN}(x) 	= 	\frac{1}{\sqrt{2 \pi }  \sigma x } \exp\left(  - \frac{ (\log x- \mu)^2 }{ 2 \sigma^2}  \right) ,	     		
\end{eqnarray} 
where, $\mu$ and $\sigma^2$ are the mean and variance of the natural logarithm of $x$. %
The parameters are $\mu= 0.01$ and $\sigma^2= 0.25$ for pairwise and three-body coupling strength in Fig.~\ref{fig:Est_Coupling} and Fig.~\ref{fig:Large_Pop}, respectively. 
With the exception of Fig. 5, the number of oscillators is $N= 12$. In Fig. 5, the number of oscillators is $N= 90$.

\subsection*{Baseline method for inferring the interaction network from time series}

We describe the baseline methods: (i) naive ordinary least squares (OLS), (ii) OLS with statistical testing, and (iii) LASSO. The performance of these methods in inferring the couplings is compared with that of the proposed method.

\subsubsection*{OLS}

We explain two methods based on OLS regression, i.e. i) naive OLS and ii) OLS with a statistical test (Benjamini-Hochberg procedure).  
Let us consider the parameters of the $i$-th oscillator: $\omega_i$, ${\bf W}^{(2)}_i$, and ${\bf W}^{(3)}_i$. These parameters are determined by minimizing the squared error given by 
\begin{equation}
	E_{\rm OLS} =		\sum_{k=0}^{L-1}	\left(	  \frac{\Delta \phi_i}{\Delta t}(kh)	 - \omega_i -  
		f^{(2)}_i( {\bf \phi}(kh); {\bf W}^{(2)}_i ) -	f^{(3)}_i( {\bf \phi}(kh); {\bf W}^{(3)}_i )	\right)^2, 
	\label{eq:LS_2}
\end{equation}
where $h$ is the sampling interval and $L$ is the total number of observations.

The parameters obtained by the OLS method cannot infer "no coupling"  (i.e. the coupling strength, $W_{ij}^{(2)}$ or $W_{ijl}^{(3)}$, is zero) when the system receives external noise. Consequently, it is necessary to distinguish "no coupling" from "weak coupling" (i.e. the coupling strength is small) by testing the null hypothesis of no coupling: $W_{ij}^{(2)}= 0$ for the pairwise coupling and $W_{ijl}^{(3)}= 0$ for the three-body coupling, respectively. 
To calculate the confidence interval, a synthetic data set without interaction, denoted as $\{ \phi_{i: W= 0}(t) \}$ ($i= 1, 2, \cdots, N$), is generated by simulating the dynamic model (\ref{eq:Kuramoto}) with the intrinsic parameters ($\omega_i, \sigma^2_i$) set to those determined by OLS regression (Eq.~\ref{eq:LS_2}) and the coupling parameters set to 0, i.e. $W^{(2)}_{ij}= 0$ and $W^{(3)}_{ijl}= 0$.
Again, the parameters, $\tilde{\omega}_i, \tilde{W}^{(2)}_{ij}, \tilde{W}^{(3)}_{ijl}$, and $\tilde{\sigma}^2_i$ ($i, j, l = 1, 2, \cdots, N$), are inferred from the synthetic data $\{ \phi_{i: W= 0}(t) \}$ ($i= 1, 2, \cdots, N$) using the OLS regression (Eq.~\ref{eq:LS_2}).  
The mean, $\tilde{\mu}^{(2)}_{W}$ and $\tilde{\mu}^{(3)}_{W}$, and the variance, $(\tilde{\sigma}^2)^{(2)}_W$ and $(\tilde{\sigma}^2)^{(3)}_W$, of the interaction strength from the null model (i.e. oscillators without interaction) are determined by the sample mean and variance: 
\begin{eqnarray}
	\tilde{\mu}^{(2)}_{W} 		&=& 	\frac{1}{N (N-1)} \sum_{i, j: i \neq j}  \tilde{W}^{(2)}_{i j},	 \quad
		(\tilde{\sigma}^2)^{(2)}_{W}	=	 	\frac{1}{N (N-1)} \sum_{i, j: i \neq j}  \left( \tilde{W}^{(2)}_{i j}- \tilde{\mu}^{(2)}_{W} \right)^2,	 \\
	\tilde{\mu}^{(3)}_{W} 		&=& 	\frac{2}{N (N-1)(N-2)} \sum_{i, j, l: i \neq j, i \neq l, l >j}  \tilde{W}^{(3)}_{i j l},	 \quad
    		(\tilde{\sigma}^2)^{(3)}_{W}	=	 	\frac{2}{N (N-1)(N-2)} \sum_{i, j, l: i \neq j, i \neq l, l >j}  \left( \tilde{W}^{(3)}_{i j l}- \tilde{\mu}^{(3)}_{W} \right)^2. 
\end{eqnarray} 
As the naive method, we detect the interaction by thresholds, $\theta^+_W$ and $\theta^-_W$, and conclude that there is an interaction if the inferred interaction strength $w_{\rm OLS}$ is larger or smaller than the threshold, $w_{\rm OLS} > \theta^+_W$ or $w_{\rm OLS}< \theta^-_W$. 
The thresholds are set by the mean and standard deviation of the interaction strength of the null model: 
$\theta^+_W= \tilde{\mu}^{(2)}_{W}+ 2 \tilde{\sigma}^{(2)}_{W}$,  $\theta^-_W= \tilde{\mu}^{(2)}_{W}- 2 \tilde{\sigma}^{(2)}_{W}$ for the pairwise interaction, and $\theta^+_W= \tilde{\mu}^{(3)}_{W}+ 2 \tilde{\sigma}^{(3)}_{W}$,  $\theta^-_W= \tilde{\mu}^{(3)}_{W}- 2 \tilde{\sigma}^{(3)}_{W}$ for the three-body interaction, respectively. 

As a more refined method, we identify the existence of couplings by using the Benjamini-Hochberg procedure~\cite{benjamini1995controlling,albert2016surrogate} to address the multiple testing problem. Specifically, we obtain the $p$-values $P^{(2)}_{ij}, P^{(3)}_{ijl}$ for the coupling strength $\hat{W}^{(2)}_{ij}, \hat{W}^{(3)}_{ijl}$ obtained from the original data by assuming that the inferred strength follows a Gaussian distribution under the null hypothesis. 
We sort the $p$-values $P^{(2)}_{ij}, P^{(3)}_{ijl}$ in ascending order, and denote them by $\tilde{P}^{(2)}_{1}, \tilde{P}^{(2)}_{2}, \cdots \tilde{P}^{(2)}_{N(N-1)}$ and $\tilde{P}^{(3)}_{1}, \tilde{P}^{(3)}_{2}, \cdots \tilde{P}^{(3)}_{N(N-1)(N-2)/2}$. 
Then, we find the largest indices $k^{(2)}$ and $k^{(3)}$ such that  $\tilde{P}^{(2)}_{k^{(2)}} \leq  \frac{k^{(2)}}{m} \alpha$ and $\tilde{P}^{(3)}_{k^{(3)}} \leq  \frac{k^{(3)}}{m} \alpha$, where $\alpha= 0.05$ is the significance level.  
Finally, we reject the null hypothesis corresponding to the $p$-values: $\tilde{P}^{(2)}_{1},  \cdots  , \tilde{P}^{(2)}_{k^{(2)}}$  and  $\tilde{P}^{(3)}_{1},  \cdots  , \tilde{P}^{(3)}_{k^{(3)}}$.  
If the null hypothesis is rejected, we detect an interaction ($W^{(2)}_{ij}$ or $W^{(3)}_{ijl}$) and its strength is estimated as $\hat{W}^{(2)}_{ij}, \hat{W}^{(3)}_{ijl}$. 
Otherwise, we infer that there is no interaction among the node pair or triple: $W^{(2)}_{ij}= 0$ or $W^{(3)}_{ijl}= 0$.

\subsubsection*{LASSO} 
We describe the inference method based on LASSO (Least Absolute Shrinkage and Selection Operator). Let us consider the interaction parameters of the $i$-th oscillator: ${\bf W}^{(2)}_i$, and ${\bf W}^{(3)}_i$. 
The parameters of the $i$-th oscillator are determined by minimizing the cost function given by 
\begin{equation}
	E_{\rm LAS} =		E_{\rm OLS}+ \alpha   
         	\left\{    \sum_{j: j \neq i}	   \left| W^{(2)}_{ij} \right| + 
		\sum_{j, l: l >j, j \neq i, l  \neq i}	   \left| W^{(3)}_{ijl} \right|    \right\},  
	\label{eq:LAS}
\end{equation}
where $E_{\rm OLS}$ is the squared error, as defined by Eq. (\ref{eq:LS_2}), and $|x|$ represents the absolute value of a real number $x$.
In contrast to the OLS method, LASSO is capable of inferring whether a coupling exists or not ($W^{(2)}_{ij}= 0$ or $W^{(3)}_{ijl}= 0$). 
The hyperparameter $\alpha$ is determined by minimizing the Bayesian Information Criterion (BIC). We use the \textit{LassoLarsIC} function from \textit{scikit-learn}~\cite{JMLR:v12:pedregosa11a}. All the interaction parameters are obtained by repeating the above procedure for each oscillator ($i= 1, 2, \cdots, N$).

\subsection*{Statistical test for identifying the interaction type}

We identify the interaction type (i.e. pairwise, three-body, or mixture) from the phase time series by applying a statistical test about the probability of the pairwise and three-body couplings. Specifically, we consider the null hypothesis that the probability of pairwise and three-body couplings are equal: $q^{(2)} = q^{(3)}$, where $q^{(2)}$ and $q^{(3)}$ represent the probabilities of the existence of the pairwise and three-body couplings, respectively. 
If the hypothesis is not rejected, we conclude that the probabilities are similar and that the interaction type is mixture. Otherwise, we conclude that the probabilities are different ($q^{(2)} \neq q^{(3)}$) and that the interaction type with the higher probability is dominant.

The procedure to identify the interaction type from the phase time series is summarized as follows: 
\begin{itemize}
	\item[1.]   Obtain the interaction parameters $W^{(2)}_{ij}$ and $W^{(3)}_{ijl}$ ($i, j, l = 1, 2, \cdots, N$) from the phase time series using an inference method (i.e., adaptive LASSO, LASSO, or OLS with a statistical test.	
	\item[2.]   Obtain the coupling probabilities $\hat{q}^{(2)}$, $\hat{q}^{(3)}$ by calculating the proportion of nonzero values in the pairwise or three-body coupling parameters: 
		\begin{eqnarray}
    			\hat{q}^{(2)}=  \frac{m^{(2)} }{n^{(2)} },	\quad	\hat{q}^{(3)}=  \frac{m^{(3)} }{n^{(3)} },		
		\end{eqnarray} 
		where, $m^{(2)}$ and $m^{(3)}$ are the numbers of non-zero pairwise and three-body coupling strengths, and $n^{(2)}= N(N-1)$ and $n^{(3)}= N(N-1)(N-2)/2$ are the total number of pairwise and three-body couplings, respectively.  
	\item[3.]   We test the null-hypothesis of $H_0: q^{(2)} = q^{(3)}$ using the $Z$-test. Under the null-hypothesis of $H: q^{(2)} = q^{(3)}$, the difference of the coupling probabilities approximately obeys the Gaussian distribution of mean $0$ and variance $\sigma^2_d= q_c (1-q_c) \left( \frac{1}{n^{(2)} } + \frac{1}{n^{(3)} } \right)$, where $q_c:= q^{(2)}= q^{(3)}$ is the overall coupling probability. Thus, we can test the null-hypothesis $H_0$ by calculating the standardized statistic: 
		\begin{eqnarray}
    			Z=  \frac{\hat{q}^{(2)}- \hat{q}^{(3)} }{\sqrt{ \hat{q}_c (1- \hat{q}_c) \left( \frac{1}{n^{(2)} } + \frac{1}{n^{(3)} } \right)} },	
		\end{eqnarray} 	 
		where,  $\hat{q}_c = \frac{m^{(2)} +m^{(3)} }{n^{(2)} +n^{(3)} }$ is an estimate of the overall coupling probability.   
	\item[4.]   If the null hypothesis $H_0$ is not rejected, we conclude that the coupling probabilities are similar and that the interaction type is mixture. 
	Otherwise, we conclude that the interaction type is the one with the higher probability, i.e., If $\hat{q}^{(2)} > \hat{q}^{(3)}$, the interaction type is inferred as pairwise, while $\hat{q}^{(3)} > \hat{q}^{(2)}$, the interaction type is inferred as three-body. Here we adopt the significance level of $\alpha= 0.05$.  
\end{itemize}

\section*{Acknowledgements}
We thank Riccardo Muolo for the constructive comments on the manuscript. This research was partially supported by the Japan Society for the Promotion of Science (JSPS) KAKENHI (Nos. JP22K11919 and JP22H00516) and JST CREST (No. JPMJCR1913) to H.N., JSPS KAKENHI (Nos. JP21K12056, JP22K18384, and JP23K27487) to H.K., and JSPS KAKENHI (Nos. JP18K11560, JP19H01133, JP21H03559, JP21H04571, and JP22H03695), JST FOREST (No. JPMJFR232O), and AMED (No. JP223fa627001) to R.K. 
\section*{Author contributions}
R.K., S.H. and H.N conceived the project. W.S. and R.K. developed the methods for inferring the network structure from time series. W.S. performed numerical simulations and analyzed data. W.S., S.H. and R.K. created the figures. R.K. and H.K supervised the project. 
R.K wrote the draft based on input from W.S., S.H. and H.N. 
All authors edited and approved the manuscript.

\section*{Competing interests}
The authors declare no competing interests.

\section*{Data availability}
The datasets and simulation codes for generating the data are available on GitHub:  
\url{https://github.com/Weiwei-Su112/Adaptive-LASSO-for-higher-order-oscillatory-systems}

\section*{Code availability}
The code for the adaptive LASSO method for inferring the network from phase time series is available on GitHub \url{https://github.com/Weiwei-Su112/Adaptive-LASSO-for-higher-order-oscillatory-systems} 

\bibliography{su_ref}
\clearpage

\begin{table}[t]
	\centering
	\caption{ {\bf Accuracy of interaction type inference.}   
    'Proposed' represents the accuracy of the proposed method. 
    'LASSO' represents the accuracy of the LASSO method.  
    'OLS' and 'OLS (Naive)' represent the accuracy of OLS with a statistical test and OLS with naive thresholding, respectively (see Methods for detail of LASSO and OLS methods). 
    {\bf Bold} letters indicate the most accurate method. The synthetic data were generated from 12 oscillators.} 
		\label{tab:Interact_Type} 
	\vspace{0.3cm}
		\begin{tabular}{|c|c|c|c|c|}
		\hline 
        Interaction type 	&  Proposed 	& 	LASSO	&	OLS 		 		&	OLS (Naive)	 \\ \hline  
		Pairwise 			& 	{\bf 100 \%}	&  	{\bf 100 \%}	&	{\bf 100 \%}		&	 {\bf 100 \%}	\\
		Three-body 		& 	{\bf 100 \%}	&  	{\bf 100 \%}	&	{\bf 100 \%}		&	 80 \%		\\
		Mixture	 		& 	{\bf 100 \%}	&  	{\bf 100 \%}	&	 85 \%		&	 85 \%		\\  \hline		
	\end{tabular}
\end{table}

\begin{table}[t]
	\centering
	\caption{ {\bf Inference of coupling probabilities.}  
    'Proposed' represents the result of the proposed method. 'LASSO' represents the result of the LASSO method. 
    'OLS' and 'OLS (Naive)' represent the result of OLS with a statistical test and OLS with naive thresholding, respectively (see Methods for detail of LASSO and OLS methods). 
    {\bf Bold} letters indicate the most accurate method. The synthetic data were generated from 12 oscillators.}		\label{tab:Interact_Prob} 
	\vspace{0.3cm}
		\begin{tabular}{|c|c|c|c|c|c|c|}
		\hline 
        \multicolumn{2}{|c|}{}						&	True		&  	Proposed 	& 	LASSO		&	OLS		&	OLS (Naive)   \\  \hline 
	\multirow{2}{*}{Pairwise}		&	$q^{(2)}$ 		& 	  10.0 \%	&      {\bf 10.4 \%}		&      	7.7 \%	   	&	 16.1 \%		&	23.1 \%  	\\	
		 	&	$q^{(3)}$ 		&	  0.0 \%	&	{\bf 0.1 \%}		&	0.7  \%		&	   0.6 \%		&	  9.2 \%	\\	\hline
	\multirow{2}{*}{Three-body }	&	$q^{(2)}$ 		& 	  0.0 \%	&      	       0.8 \%		&     {\bf 0.0 \%}	   	&	   0.1  \%		&	  5.1 \%	\\	
		 	&	$q^{(3)}$ 		&	  10.0 \%	&	{\bf   10.1 \%}		&	10.8  \%		&	   8.3 \%		&	  13.7 \%	\\	\hline
	\multirow{2}{*}{Mixture }		&	$q^{(2)}$ 		& 	  4.9 \%	&      	 5.6 \%		&      	{\bf 4.3 \%}   	&	   6.1  \%		&	12.2 \%	\\	
		 	&	$q^{(3)}$ 		&	  4.9 \%	&	{\bf   4.9 \%}		&	5.4  \%		&	   4.2 \%		&	  10.2 	\%  \\	\hline
	\end{tabular}
\end{table}

\begin{table}[t]
	\centering
	\caption{ {\bf Accuracy of coupling network inference.} 
    The accuracy was evaluated by MCC. 
    {\bf Bold} letters indicate the best method.} 		\label{tab:MCC_small_LN} 
	\vspace{0.3cm}
		\begin{tabular}{|c|c|c|c|}
		\hline         
        Interaction type 	&  Proposed 	& 	LASSO	&	OLS 		 \\ \hline		            
		Pairwise 		& 	{\bf 0.92}		&  	0.59		&	0.60	 	 \\
		Three-body 		& 	{\bf 0.85}		&  	0.83		&	0.81		 \\
		Mixture	 		& 	{\bf 0.85}		&  	0.83		&	0.78		 \\  \hline	
	\end{tabular}
\end{table}

\begin{table}[t]
	\centering
	\caption{ {\bf Accuracy of coupling strength inference.}   
    The accuracy was evaluated by mean squared error. {\bf Bold} letters indicate the best method.} 		\label{tab:MSE_small_LN} 
	\vspace{0.3cm}
		\begin{tabular}{|c|c|c|c|}		\hline 
        Interaction type 	&  Proposed	 & 	LASSO	&	OLS 		 \\ \hline
            Pairwise 		& 	${\bf 1.5 \times 10^{-5}}$		&  	$7.1 \times 10^{-5}$		&	$3.0 \times 10^{-4}$	 	 \\
		Three-body 		& 	${\bf 2.4 \times 10^{-5}}$		&  	$4.0 \times 10^{-5}$		&	$1.7 \times 10^{-4}$	\\
		Mixture	 		& 	${\bf 1.6 \times 10^{-5}}$		&  	$3.6 \times 10^{-5}$		&	$2.0 \times 10^{-4}$	 	 \\  \hline	
	\end{tabular}
 \end{table}

\begin{table}[t]
	\centering
	\caption{    
    {\bf Accuracy of coupling inference of the human brain network.}  
    The accuracy is evaluated by MCC. 
    {\bf Bold} letters indicate the best method.  
    }\label{tab:MCC_HB_LN} 
	\vspace{0.3cm}
		\begin{tabular}{|c|c|c|c|}
		\hline         
            Subjects ID 	&  Proposed 	& 	LASSO	    &	OLS 		 \\ \hline		
		S001 		& 	{\bf 0.659}		&  	0.353	&  0.062	 	 \\
		S026 		& 	{\bf 0.630}		&  	0.335	&	0.066		 \\
        S038 		& 	{\bf 0.694}		&  	0.345	&	0.132		 \\
		S062	 	& 	{\bf 0.618}		&  	0.513	&	0.061		 \\  \hline	
	\end{tabular}
\end{table}

\end{document}


\title{Supplementary material for Identification of higher order interactions in phase oscillator networks from time series}
\date{\today}
\maketitle

\section*{Supplementary Note 1: Relationship between coupling strength and inference accuracy} 
We examined the relationship between coupling strength and inference accuracy (Supplementary Fig.~1). Similar to Fig. 3, the inference performance was evaluated using the Matthews correlation coefficient (MCC).
As expected, a stronger coupling strength leads to higher accuracy. For a pairwise interaction ($q^{(2)} = 0.05$), the OLS method fails to accurately infer the interaction network, even under strong coupling conditions. 
Overall, the proposed method consistently outperforms existing approaches (LASSO and OLS) in inferring network topology from phase time series.

\vspace{1cm}
\begin{figure}[h]
\begin{center}
    \label{fig:MCC_vs_K}
    \includegraphics[scale=.75]{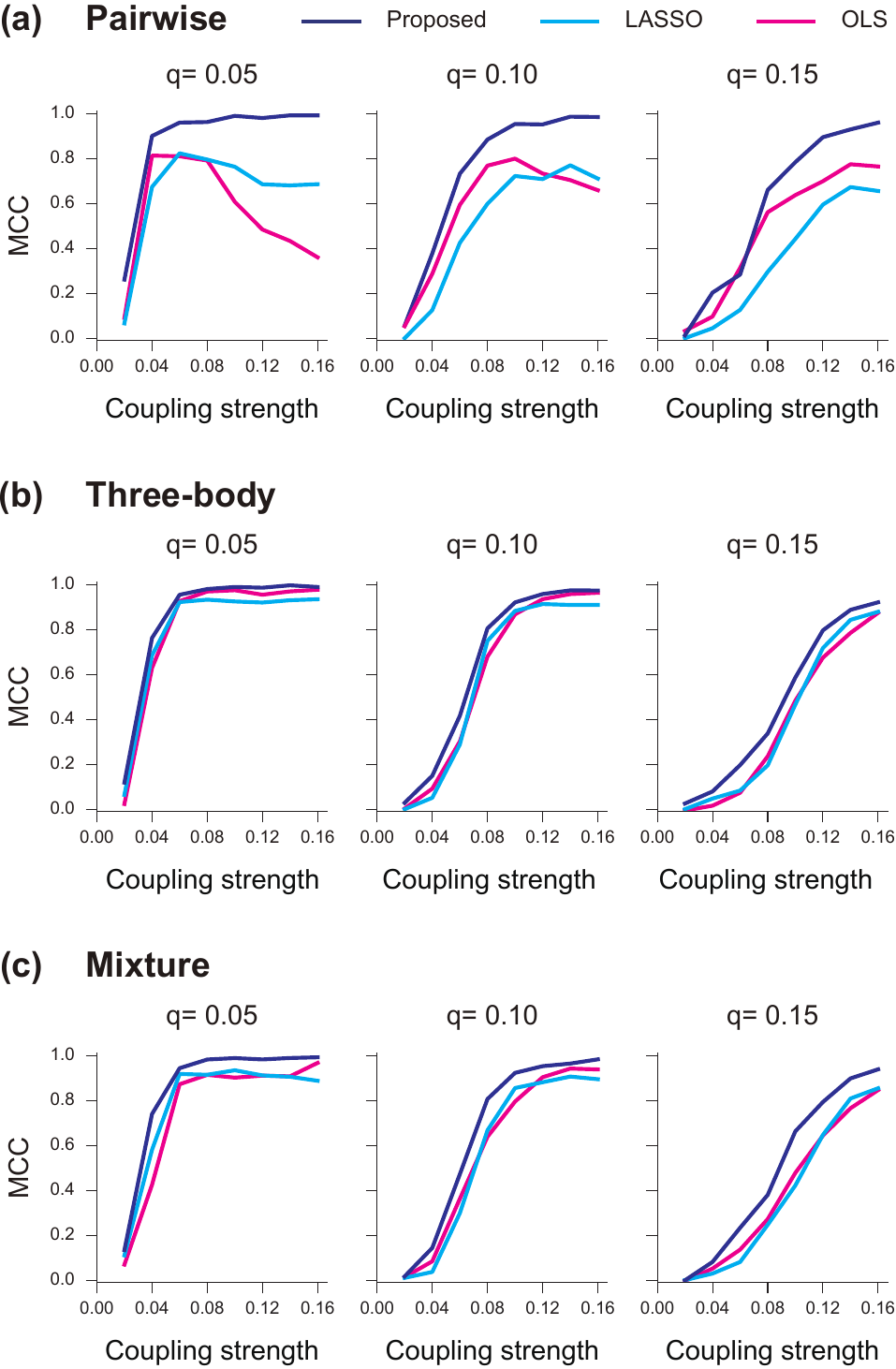} 
    \ \\
    \textbf{Supplementary Figure 1}. 
    \textbf{Effect of coupling strength on inference accuracy.} 
    Performance of coupling inference (MCC) was calculated for the three types of interaction: 
    (a) Pairwise ($q^{(2)}= q$, $q^{(3)}= 0$), 
    (b) Three-body ($q^{(2)}= 0$, $q^{(3)}= q$), and 
    (c) Mixture of pairwise and three-body interaction ($q^{(2)}= q/2$, $q^{(3)}= q/2$).  The parameter of the coupling probability $q$ was set to 0.05, 0.10 and 0.15 on the left, centre and right panel, respectively. 
\end{center}
\end{figure}
\clearpage

\section*{Supplementary Note 2: Coupled oscillators with phase-dependent amplitude} 
We consider weakly coupled oscillators with phase-dependent amplitude, which could be a reasonable model for real-world oscillations~\cite{bick2024higher}. 
Assuming the symmetry in three-body interaction (e.g., $W^{(3)}_{i j l}= W^{(3)}_{i l j}$), 
a phase model is derived from the weakly coupled oscillators by using the second-order phase reduction~\cite{bick2024higher} 
\begin{equation}    	
    \frac{d\phi_i}{d t}=	\omega_i +  f^{(2,1)}_i \left( \vec{\phi}; {\bf W}^{(2,1)}_i \right) + f^{(2,2)}_i \left( \vec{\phi}; {\bf W}^{(2,2)}_i \right) 
    +  f^{(3, 1)}_i \left( \vec{\phi}; {\bf W}^{(3, 1)}_i \right) + f^{(3, 2)}_i \left( \vec{\phi}; {\bf W}^{(3, 2)}_i \right) + \sigma_i \xi_i(t), \quad  \ 
    \label{eq:SN_Osc_gen_3-body}
\end{equation}  
where 
\begin{eqnarray}		
    f^{(2,1)}_i \left( \vec{\phi}; {\bf W}^{(2,1)}_i \right)  &=& 	 \frac{1}{ q^{(2,1)} N} \sum_{j: j \neq i}^N  W^{(2,1)}_{i j} \sin \left( \phi_j-\phi_i + \psi  \right),	\label{eq:SN_Intgen_pairwise_1}  \\
    f^{(2,2)}_i \left( \vec{\phi}; {\bf W}^{(2,2)}_i \right)  &=& 	 \frac{1}{ q^{(2,2)} N} \sum_{j: j \neq i}^N  W^{(2,2)}_{i j} \sin \left( \phi_j-\phi_i + 2 \psi \right),	\label{eq:SN_Intgen_pairwise_2}  \\
    f^{(3, 1)}_i \left( \vec{\phi}; {\bf W}^{(3, 1)}_i \right)  &=&	 \frac{1}{ q^{(3,1)} N^2} \sum_{j, l: l>j}^N  W^{(3,1)}_{i j l}  \{  \sin \left(2 \phi_j-\phi_l-\phi_i \right)
    +\sin \left(2 \phi_l-\phi_i-\phi_j \right)  \},	 	\label{eq:SN_Intgen_3body_1}  \\
    f^{(3, 2)}_i \left( \vec{\phi}; {\bf W}^{(3, 2)}_i \right)  &=&	 \frac{1}{ q^{(3,2)} N^2} \sum_{j, l: l>j}^N  W^{(3,2)}_{i j l}  \sin \left( \phi_j+\phi_l-2 \phi_i + 2 \psi \right),    \label{eq:SN_Intgen_3body_2}
\end{eqnarray} 
where $\vec{\phi}= (\phi_1, \cdots, \phi_N)$ is the phase of oscillators, and  
${\bf W}^{(2,1)}_i= (W^{(2,1)}_{i1}, \cdots,  W^{(2,1)}_{iN}) $, 
${\bf W}^{(2,2)}_i= (W^{(2,2)}_{i1}, \cdots,  W^{(2,2)}_{iN}) $,
${\bf W}^{(3, 1)}_i= (W^{(3, 1)}_{i12}, \cdots, W^{(3, 1)}_{i1N}, W^{(3, 1)}_{i23}, \cdots, W^{(3, 1)}_{i N-1 N})$, and 
${\bf W}^{(3, 2)}_i= (W^{(3, 2)}_{i12}, \cdots, W^{(3, 2)}_{i1N}, W^{(3, 2)}_{i23}, \cdots, W^{(3, 1)}_{i N-1 N})$
are the vectors of pairwise and three-body coupling strength, respectively. Here we assume the self-couplings are zero, e.g. $W^{(2,1)}_{ii}= 0$ and $W^{(3,1)}_{ijk}= 0$ for $i=j$ or $i=k$. 
In this study, we assume a special case of $\psi = 0$, for the sake of simplicity, and then obtain
\begin{equation}    	
    \frac{d\phi_i}{d t}=	\omega_i +  f^{(2)}_i \left( \vec{\phi}; {\bf W}^{(2)}_i \right) 
    +  f^{(3, 1)}_i \left( \vec{\phi}; {\bf W}^{(3, 1)}_i \right) + f^{(3, 2)}_i \left( \vec{\phi}; {\bf W}^{(3, 2)}_i \right) + \sigma_i \xi_i(t) ,
    \label{eq:SN2_Osc_gen_3-body}
\end{equation}  
where 
\begin{eqnarray}		
    f^{(2)}_i \left( \vec{\phi}; {\bf W}^{(2)}_i \right)  &=& 	 \frac{1}{ q^{(2)} N} \sum_{j: j \neq i}^N  W^{(2)}_{i j} \sin \left( \phi_j-\phi_i \right),	\label{eq:SN2_Intgen_pairwise}  \\   
    f^{(3, 1)}_i \left( \vec{\phi}; {\bf W}^{(3, 1)}_i \right)  &=&	 \frac{1}{ q^{(3,1)} N^2} \sum_{j, l: l>j}^N  W^{(3,1)}_{i j l}  \{  \sin \left(2 \phi_j-\phi_l-\phi_i \right)
    +\sin \left(2 \phi_l-\phi_i-\phi_j \right)  \},	 	\label{eq:SN2_Intgen_3body_1}  \\
    f^{(3, 2)}_i \left( \vec{\phi}; {\bf W}^{(3, 2)}_i \right)  &=&	 \frac{1}{ q^{(3,2)} N^2} \sum_{j, l: l>j}^N  W^{(3,2)}_{i j l}  \sin \left( \phi_j+\phi_l-2 \phi_i \right).    \label{eq:SN2_Intgen_3body_2}
\end{eqnarray} 
As with the Kuramoto model with three-body interactions (see Results for details), we infer the coupling strength by using the adaptive Lasso. 
First, a preliminary estimate of the parameters of the $i$-th oscillator ($\omega_i$, ${\bf W}^{(2)}_i$, ${\bf W}^{(3,1)}_i$, and ${\bf W}^{(3,2)}_i$) are obtained by OLS~\cite{malizia2024reconstructing}. 
These parameters are determined through minimizing the squared error given by
\begin{equation}
    E_{\rm OLS} =		\sum_{k=0}^{L-1}	\left(	  \frac{\Delta \phi_i}{\Delta t}(kh)	 - \omega_i -  
	f^{(2)}_i( {\bf \phi}(kh); {\bf W}^{(2)}_i ) -	f^{(3,1)}_i( {\bf \phi}(kh); {\bf W}^{(3,1)}_i ) 
    -	f^{(3,2)}_i( {\bf \phi}(kh); {\bf W}^{(3,2)}_i )	\right)^2, 
    \label{eq:SN1_OLS}
\end{equation}
where $h$ is the sampling interval. 
The estimate is then refined through minimizing the cost function given by 
\begin{equation}
    E_{\rm AL} =  E_{\rm OLS}+ \alpha	    
    \left\{    \sum_{j}	     \left|  c^{(2)}_{ij}  W^{(2)}_{ij}  \right| + 
    \sum_{j, l: l >j}     \left|  c^{(3,1)}_{ijl} W^{(3,1)}_{ijl}   \right| 
    + \left|  c^{(3,2)}_{ijl} W^{(3,2)}_{ijl}   \right|
    \right\},   		
\end{equation}
where $|x|$ represents the absolute value of a real number $x$, and $c^{(2)}_{ij}$, $c^{(3,1)}_{ijl}$, and $c^{(3,2)}_{ijl}$ are hyperparameters, which are set as follows: 
$c^{(2)}_{ij}= 1/ \tilde{W}^{(2)}_{ij}$,
$c^{(3,1)}_{ijl}= 1/ \tilde{W}^{(3,1)}_{ijl}$, and $c^{(3,2)}_{ijl}= 1/ \tilde{W}^{(3,2)}_{ijl}$, 
where $\tilde{W}^{(2)}_{ij}$, $\tilde{W}^{(3,1)}_{ijl}$, and $\tilde{W}^{(3,2)}_{ijl}$ are the preliminary estimates obtained by OLS (Eq. 25). 
For a fixed $\alpha$, the parameters ($W^{(2)}_{ij}$, $W^{(3,1)}_{ijl}$, and $W^{(3,2)}_{ijl}$) are fitted by using the Least Angle Regression (LARS)~\cite{efron2004least}. 
A hyperparameter $\alpha$ is determined by minimizing the Bayesian Information Criterion (BIC), which is performed by using \textit{LassoLarsIC} function from \textit{scikit-learn} \cite{JMLR:v12:pedregosa11a}. 
\clearpage

\section*{Supplementary Note 3: Kuramoto model with four-body interaction}
To examine the feasibility for the higher order interaction (more than 3-body interaction), we consider the Kuramoto model with four-body interaction: 
\begin{equation}    	
	\frac{d\phi_i}{d t}=	\omega_i +  f^{(2)}_i \left( \vec{\phi}; {\bf W}^{(2)}_i \right) +  f^{(3)}_i \left( \vec{\phi}; {\bf W}^{(3)}_i \right) +  f^{(4)}_i \left( \vec{\phi}; {\bf W}^{(4)}_i \right) + \sigma_i \xi_i(t) ,
	\label{eq:SI-Kuramoto_4-body}
\end{equation} 
where 
\begin{eqnarray}		
    f^{(2)}_i \left( \vec{\phi}; {\bf W}^{(2)}_i \right)  &=& 	 \frac{1}{ q^{(2)} N} \sum_{j: j \neq i}^N  W^{(2)}_{i j} \sin \left( \phi_j-\phi_i \right),	\label{eq:SI-Intgen_pairwise}  \\    
    f^{(3)}_i \left( \vec{\phi}; {\bf W}^{(3)}_i \right)  &=&	 \frac{1}{ q^{(3)} N^2} \sum_{j, l: l>j}^N  W^{(3)}_{i j l}  \{  \sin \left(2 \phi_j-\phi_l-\phi_i \right)
    +\sin \left(2 \phi_l-\phi_i-\phi_j \right)  \},	 	\label{eq:SI-Intgen_3body}  \\
    f^{(4)}_i \left( \vec{\phi}; {\bf W}^{(4)}_i \right)  &=&	 \frac{1}{ q^{(4)} N^3} \sum_{j, l, m: m>l>j}^N  W^{(4)}_{i j l}  \{\sin \left( \phi_j+\phi_l- \phi_m - \phi_i \right) + \sin \left( \phi_l+\phi_m- \phi_j - \phi_i \right)  \nonumber  \\
    && 
    \quad \quad \quad \quad \quad \quad 
    \quad \quad \quad \quad 
    +\sin \left(\phi_m+\phi_j- \phi_l - \phi_i \right)\}, 
     \label{eq:SI-Intgen_4body}
\end{eqnarray} 
where $\vec{\phi}= (\phi_1, \cdots, \phi_N)$ is the phase of oscillators, and  
${\bf W}^{(2)}_i= (W^{(2)}_{i1}, \cdots,  W^{(2)}_{iN}) $, ${\bf W}^{(3)}_i= (W^{(3)}_{i12}, \cdots, W^{(3)}_{i1N}, W^{(3)}_{i23}, \cdots, W^{(3)}_{i N-1 N})$, and 
${\bf W}^{(4)}_i= (W^{(4)}_{i123}, \cdots, W^{(4)}_{i12N}, W^{(4)}_{i134}, \cdots, W^{(4)}_{i 1\ N-1\ N}, W^{(4)}_{i 2\ N-1\ N}, \cdots, W^{(4)}_{i\ N-2\ N-1\ N})$
are the vectors of pairwise, three-body, and four-body coupling strengths, respectively. 
We assume the self-couplings are zero, e.g., $W^{(2)}_{ii}= 0$, $W^{(3)}_{ijl}= 0$ for $i=j$ or $i=l$, and $W^{(4)}_{ijkm}= 0$ for $i=j$, $i=l$, or $i=m$.

As with the Kuramoto model with three-body interactions (see Results for details), we infer the coupling strength by using the adaptive Lasso. 
First, a preliminary estimate of the parameters of the $i$-th oscillator ($\omega_i$, ${\bf W}^{(2)}_i$, ${\bf W}^{(3)}_i$, and ${\bf W}^{(4)}_i$) is obtained by OLS~\cite{malizia2024reconstructing}. 
These parameters are determined through minimizing the squared error given by
\begin{equation}
    E_{\rm OLS} =		\sum_{k=0}^{L-1}	\left(	  \frac{\Delta \phi_i}{\Delta t}(kh)	 - \omega_i -  
	f^{(2)}_i( {\bf \phi}(kh); {\bf W}^{(2)}_i ) -	f^{(3)}_i( {\bf \phi}(kh); {\bf W}^{(3)}_i ) 
    -	f^{(4)}_i( {\bf \phi}(kh); {\bf W}^{(4)}_i )	\right)^2, 
    \label{eq:SN2_OLS}
\end{equation}
where $h$ is the sampling interval. 
The estimate is then refined by minimizing the cost function given by 
\begin{equation}
    E_{\rm AL} =  E_{\rm OLS}+ \alpha	    
    \left\{    
        \sum_{j}	     \left|  c^{(2)}_{ij}  W^{(2)}_{ij}  \right| + 
        \sum_{j, l: l >j}     \left|  c^{(3)}_{ijl} W^{(3)}_{ijl}   \right| + 
        \sum_{j, l, m: m> l >j}  \left|  c^{(4)}_{ijlm} W^{(4)}_{ijlm}   \right|
    \right\},   
\end{equation}
where $|x|$ represents the absolute value of a real number $x$, and $c^{(2)}_{ij}$, $c^{(3)}_{ijl}$, and $c^{(4)}_{ijl}$ are hyperparameters, which are set as follows: 
$c^{(2)}_{ij}= 1/ \tilde{W}^{(2)}_{ij}$,
$c^{(3)}_{ijl}= 1/ \tilde{W}^{(3)}_{ijl}$, and $c^{(4)}_{ijlm}= 1/ \tilde{W}^{(4)}_{ijlm}$, 
where $\tilde{W}^{(2)}_{ij}$, $\tilde{W}^{(3)}_{ijl}$, and $\tilde{W}^{(4)}_{ijlm}$ are the preliminary estimates obtained by OLS (Supplementary Eq.~\ref{eq:SN2_OLS}). 
For a fixed $\alpha$, the parameters ($W^{(2)}_{ij}$, $W^{(3)}_{ijl}$, and $W^{(4)}_{ijlm}$) are fitted by using the Least Angle Regression (LARS)~\cite{efron2004least}. 
A hyperparameter $\alpha$ is determined by minimizing the Bayesian Information Criterion (BIC), which is performed using \textit{LassoLarsIC} function from \textit{scikit-learn} \cite{JMLR:v12:pedregosa11a}. 

\bibliography{su_ref}
